\documentclass[preprint, twocolumn]{aastex62}

\usepackage{color}

\newcommand{\mum}{$\mu$m}
\newcommand{\HII}{\textrm{H~{\textsc{ii}}}}
\usepackage{graphicx,epstopdf}
\usepackage{multirow}
\usepackage{hyperref}
\usepackage{natbib}
\usepackage{array}
\usepackage{verbatim}
\usepackage{float}
\usepackage{epsfig}

\usepackage{subfigure}

\graphicspath{ {images/} }

\shorttitle{}
\shortauthors{Knight et al.}
\begin{document}

\title{Tracing PAH Size in Prominent Nearby Mid-Infrared Environments}

\author{C.~Knight}
\affiliation{Department of Physics and Astronomy, The University of Western Ontario, London, ON N6A 3K7, Canada}
\email{cknigh24@uwo.ca}

\author[0000-0002-2541-1602]{E.~Peeters}
\affiliation{Department of Physics and Astronomy, The University of Western Ontario, London, ON N6A 3K7, Canada}
\affiliation{Institute for Earth and Space Exploration, University of Western Ontario, London, ON, N6A 3K7, Canada}
\affiliation{Carl Sagan Center, SETI Institute, 189 N. Bernardo Avenue, Suite 100, Mountain View, CA 94043, USA}

\author{D.~J.~Stock}
\affiliation{Department of Physics and Astronomy, The University of Western Ontario, London, ON N6A 3K7, Canada}

\author[0000-0002-9123-0068]{W.~D.~Vacca}
\affiliation{SOFIA-USRA, NASA Ames Research Center, MS N232-12, Moffett Field, CA 94035-1000, USA}

\author[0000-0003-0306-0028]{A.~G.~G.~M.~Tielens}
\affiliation{Leiden Observatory, PO Box 9513, 2300 RA Leiden, The Netherlands}
\affiliation{Department of Astronomy, University of Maryland, MD 20742, USA}


\begin{abstract}

We present observations from the First Light Infrared TEst CAMera (FLITECAM) on--board the Stratospheric Observatory for Infrared Astronomy (SOFIA), the  {\it Spitzer} Infrared Array Camera (IRAC) and the {\it Spitzer} Infrared Spectrograph (IRS) SH mode in three well-known Photodissocation Regions (PDRs), the reflection nebulae (RNe) NGC~7023 and NGC~2023 and to the southeast of the Orion Bar, which are well suited to probe emission from Polycyclic Aromatic Hydrocarbon molecules (PAHs). We investigate the spatial behaviour of  the FLITECAM~3.3~$\mu$m filter as a proxy for the 3.3~$\mu$m PAH band, the integrated 11.2~$\mu$m PAH band, and the IRAC~8.0~$\mu$m  filter as a proxy for the sum of the 7.7 and 8.6~$\mu$m PAH bands. The resulting ratios of 11.2/3.3 and IRAC~8.0/11.2 provide an approximate measure of the average PAH size and PAH ionization respectively. In both RNe, we find that the relative PAH ionization and the average PAH size increases with decreasing distance to the illuminating source. The average PAH sizes derived for NGC~2023 are greater than those found for NGC~7023 at all points. Both results indicate that PAH size is dependent on the radiation field intensity. These results provide additional evidence of a rich carbon-based chemistry driven by the photo-chemical evolution of the omnipresent PAH molecules within the interstellar medium. In contrast, we did not detect a significant variation in the average PAH size found in the region southeast of the Orion Bar and report a peculiar PAH ionization radial profile.

\end{abstract}

\section{Introduction}
\label{intro}

The mid-infrared (MIR) spectra of many astronomical sources are dominated by emission features at  3.3, 6.2, 7.7, 8.6, 11.2, and 12.7 $\mu$m, attributed to the infrared (IR) fluorescence of Polycyclic Aromatic Hydrocarbon molecules \citep[PAHs, e.g.][]{all85,leg84}. It has been well established that PAH molecules with sizes from 50--100~carbon atoms are the carriers of these bands \citep[e.g.][]{all89,pug89}. In recent works, different authors have adopted various monikers to refer to these features such as the aromatic emission features \citep[AEFs, e.g.][]{wer04b}, the aromatic infrared bands \citep[AIB, e.g.][]{cro16} or simply referred to them as the PAH emission features \citep[e.g.][]{boe12,pee12,pee17}. Other related species are also considered to be carriers such as polycyclic aromatic nitrogen heterocycles \citep[PANHs,][]{hud05,bau08} or PAHs with functional groups attached  \citep[e.g.][]{job96,slo97,pil15,mal16,sha19}. These PAH emission features have been found in wide variety of Galactic and extragalactic sources such as \HII\, regions, reflection nebulae (RNe), post-AGB stars, planetary nebulae (PNe), the diffuse interstellar medium, and galaxies  \citep[e.g.][]{sell96,hon01,pee02,ver01}.

The PAH emission features show variations in intensities, peak positions, and band profiles \citep[e.g.][]{bre89a,hon01,pee02,gal08}. Relative intensity variations reveal how these bands relate to each other and to the intrinsic properties of these species. The 6.2, 7.7, 8.6~$\mu$m bands are well correlated over a wide range of environments \citep{job96,gal08}. Likewise, the 3.3 and 11.2~$\mu$m bands are well correlated \citep{rus77,hon01} but do not strongly correlate with the 6--9 $\mu$m PAH bands. Laboratory and quantum chemical studies show that the most significant driver of these intensity variations is the PAH charge state: 6.2, 7.7, 8.6~$\mu$m bands are prominent in ionized PAHs, whereas the 3.3 and 11.2~$\mu$m bands are strongest in neutral PAHs \citep[e.g.][]{all89,bau08}. This in turn is strongly dependent upon the UV radiation field of nearby stellar sources incident upon the environment in which these PAHs reside \citep[as well as local gas density and temperature, e.g.][]{tie05,gal08,boe13,sto16,sto17,pee17}.

The MIR spectra can be strongly influenced not only by ionization, but also by changes in the chemical structure of these species. In particular, changes in PAH size can be traced by relative intensity variations within an extended region  \citep[e.g.][]{mor12,cro16}. We investigate the PAH size distribution through the measurement of the 11.2/3.3 PAH intensity ratio in the photodissociation regions (PDRs) adjacent to three astronomical sources: two reflection nebulae (RNe): NGC~2023 and NGC~7023; and the region southeast from the Orion Bar. RNe are ideal test beds for the study of the PAH emission bands as they have strong MIR emission and have a well defined structure without contamination from atomic lines \citep[i.e.][]{sell83,rap05,ber07,boe13, boe16}, whereas the Orion Bar has long been considered the prototypical PDR \citep{tie85b}, i.e. regions where far-ultraviolet (FUV) photons with energies between 6~eV and 13.6~eV control the physics and chemistry of the gas. 

In this paper, we relate the PAH sizes in each source considered (NGC~2023, NGC~7023, and the region  southeast of the Orion Bar) to other properties of these species as well as the physical characteristics of their PDRs. Section \ref{sources} details the properties of each of the sources studied here. In Section \ref{obs}, the observations used in this study are detailed while in Section \ref{data}, the data reduction performed is discussed. Section \ref{results} presents the results of our analysis in the form of spatial maps of emission feature ratios and line projections of these ratios extending from the illuminating sources. These results are discussed in Section \ref{discussion} and conclusions are given in Section \ref{conclusion}.

\section{Astronomical Sources}
\label{sources}

For our analysis, we selected three well known sources that show prominent PAH emission features. We briefly summarize the general morphology and properties of these sources relevant to this study below.

\begin{figure*}[htbp]
\begin{center}
\includegraphics[width =8.cm]{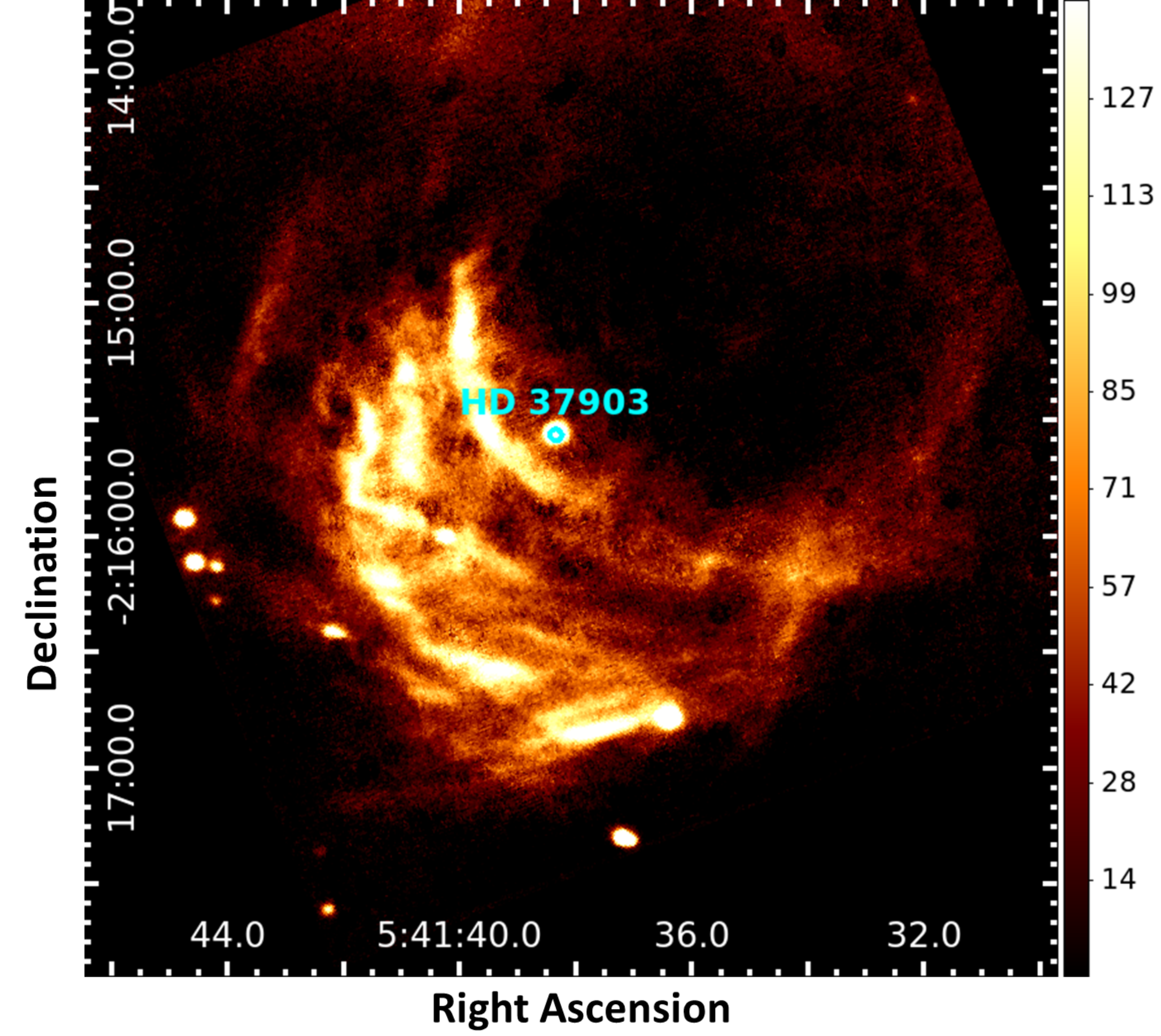}
\includegraphics[width=8. cm]{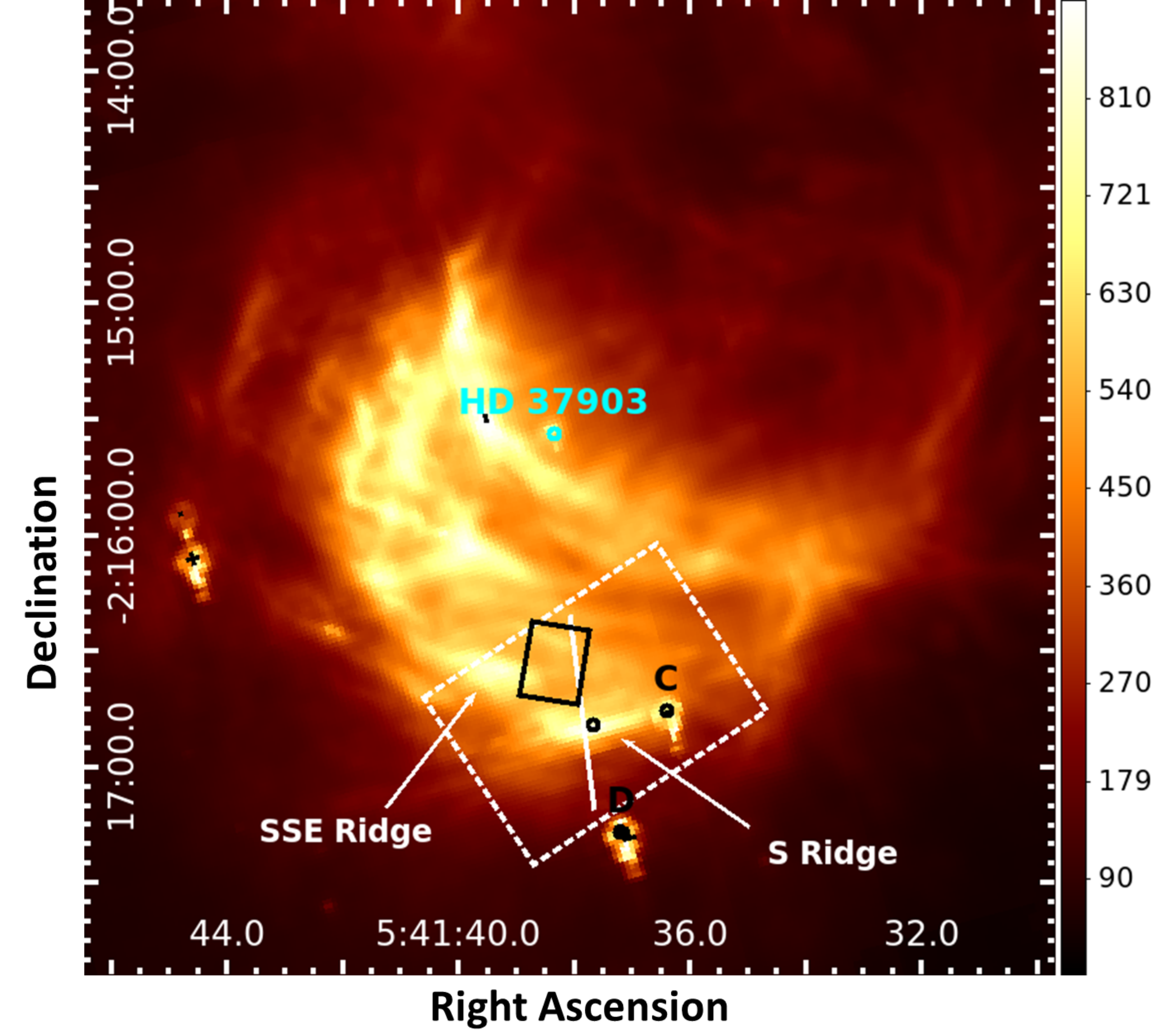}
\includegraphics[width=8. cm]{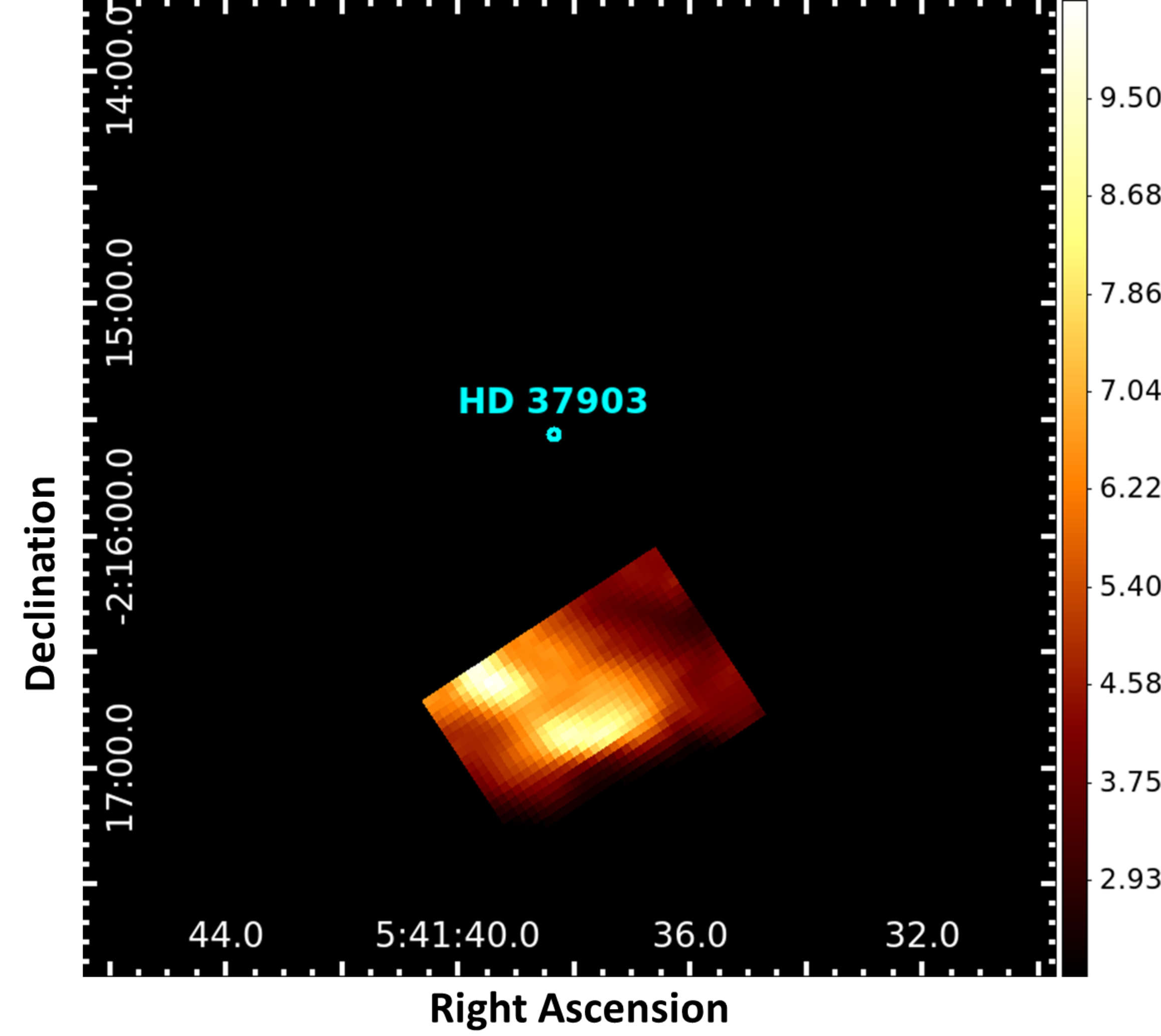}
\end{center}
\caption{FLITECAM~3.3~$\mu$m (top left), IRAC~8.0~$\mu$m (top right), and the 11.2~$\mu$m PAH band extracted from IRS~SH (bottom center) observations of NGC~2023. The illuminating star, HD~37903, is indicated by a cyan circle and the young stellar objects, C and D, are indicated by black circles, the south IRS~SH FOV by a white--dashed rectangle, the ISO-SWS FOV by a black rectangle. The crosscut used is shown as a white line, at the same position of crosscut used by \citet[][]{pee17}. The south ridge and south-southeast ridge as referred to by \cite{pee17} are labelled by S~Ridge and SSE~Ridge respectively. A 2MASS point source embedded within the S Ridge is indicated by a black circle (see Section \ref{discussion}).  The units are surface brightness (MJy/sr) for FLITECAM~3.3~$\mu$m and IRAC~8.0~$\mu$m images and integrated fluxes ($\times$~10$^{-6}$~W~m$^{-2}$~sr$^{-1}$) for the 11.2~$\mu$m image.}
\label{ngc2023_images}
\end{figure*}

\paragraph{NGC~2023} NGC~2023 is a well-known RN illuminated by the B1.5V type star, HD~37903, and located at a distance of 403~$\pm$~4 pc \citep{kou18}.  Its MIR emission has been the focus of many studies over the past 40 years \citep[e.g.][]{sell84,gat87,abe02,pee12,pee17}. It shows the same morphology at a wide range of wavelengths: a limb-darkened shell or a `bowl' shaped PDR surrounding the interior cavity (Figure \ref{ngc2023_images}). We can see into the various layers of the PDR along with some diffuse filamentary structure throughout the RN. PDR models of the cavity have shown a FUV radiation field strength of $\sim$ 4000 times that of the average interstellar field \citep[position H5,][]{ste97} as well as gas densities ranging from 10$^{4}$ to 10$^{5}$ cm$^{-3}$ \citep[e.g.][]{bur98,wyr00,san15}. We focus our study of this RN to the south of HD~37903, where the presence of multiple MIR emission peaks are suggestive of significant PAH emission along the PDR boundaries and are associated with the densest gas structures in the region. These peaks are labelled as in \cite{pee17} as the S ridge, the SSE ridge, the SE ridge and the S' ridge.

\begin{figure*}[htbp]
\begin{center}
\includegraphics[clip,trim =.cm 0.0cm .0cm 0.0cm,width=7.5 cm]{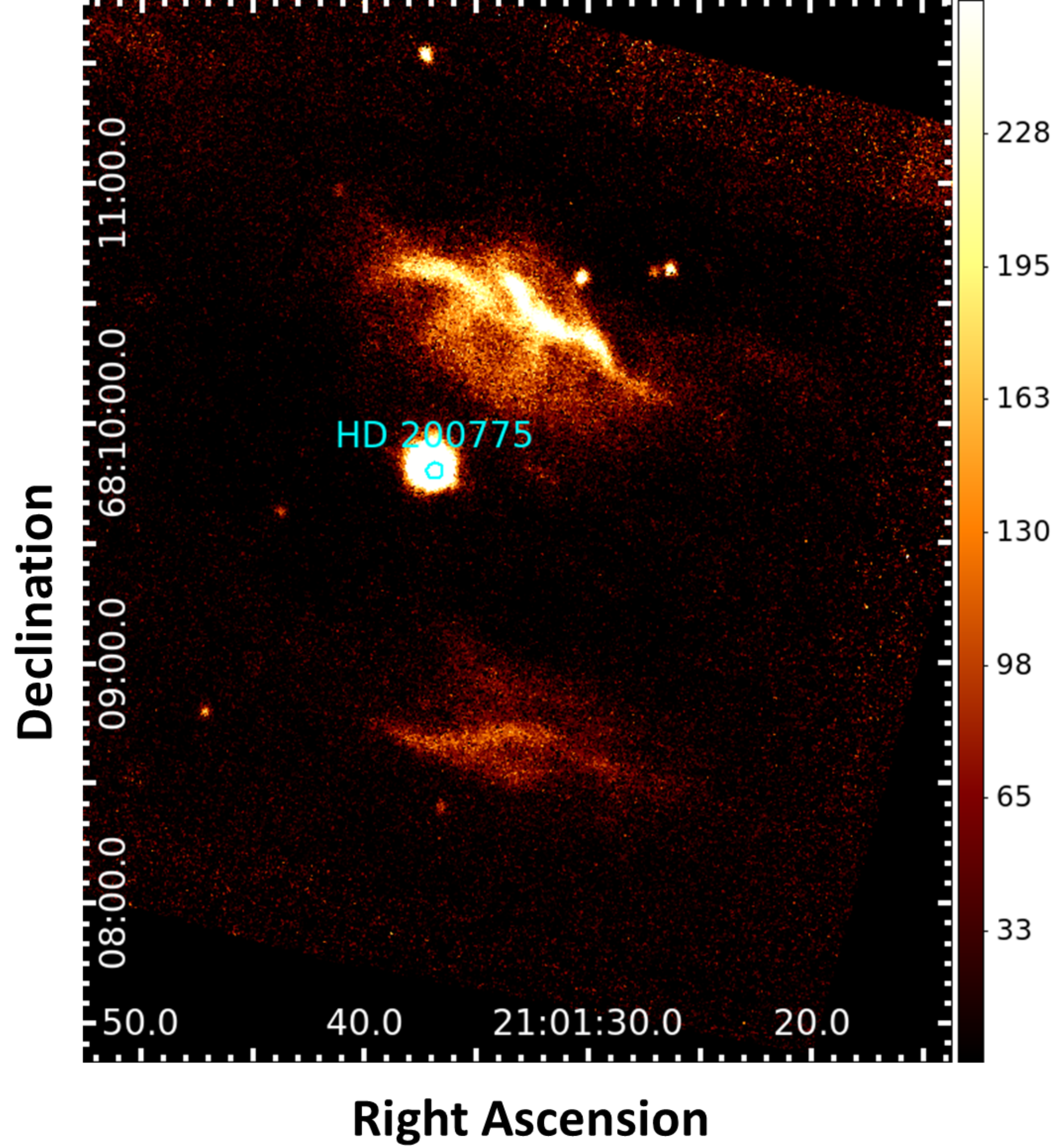}
\includegraphics[clip,trim =.cm 0.0cm .0cm 0.0cm,width=7.5 cm]{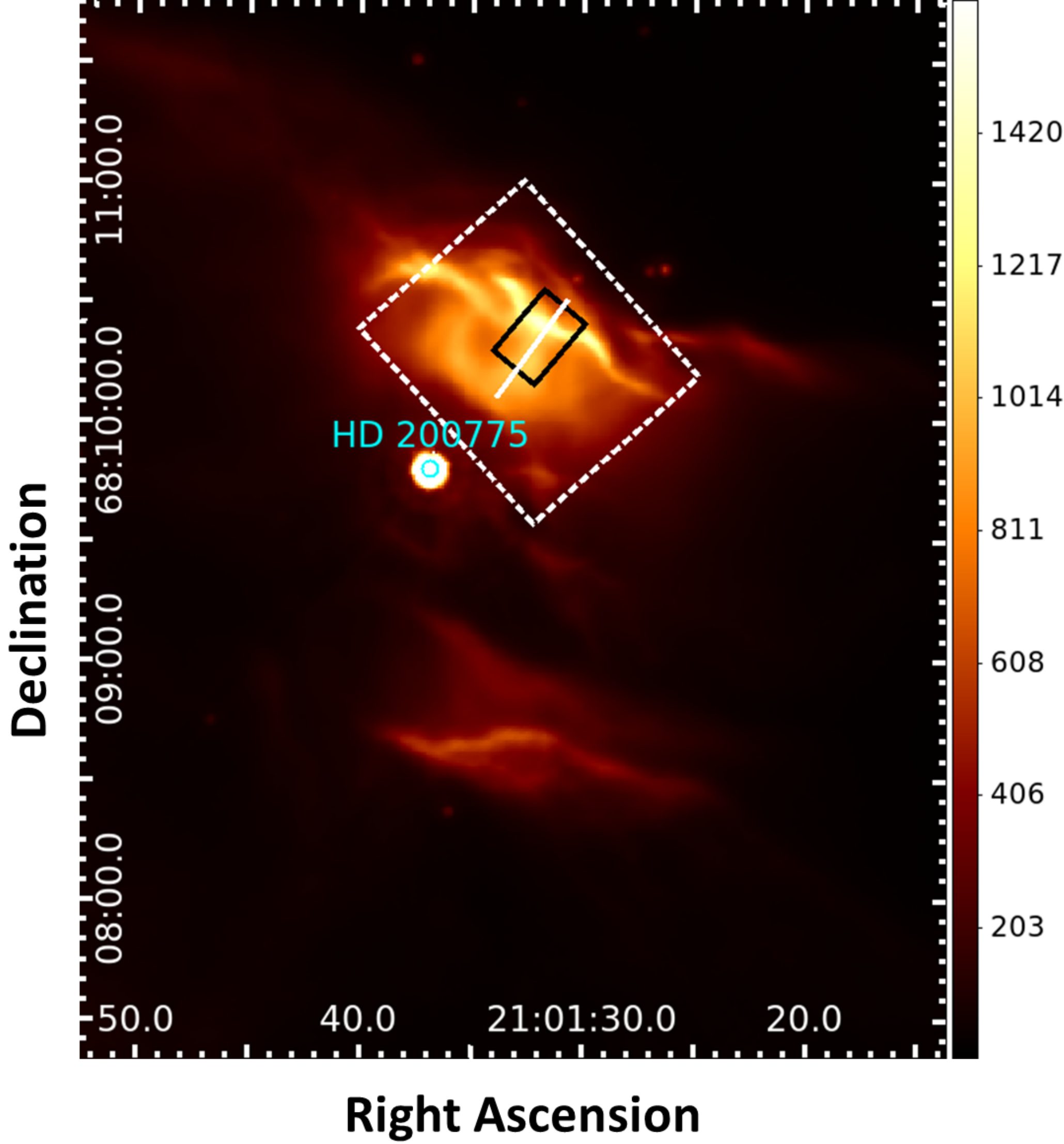}
\includegraphics[width=7.5 cm]{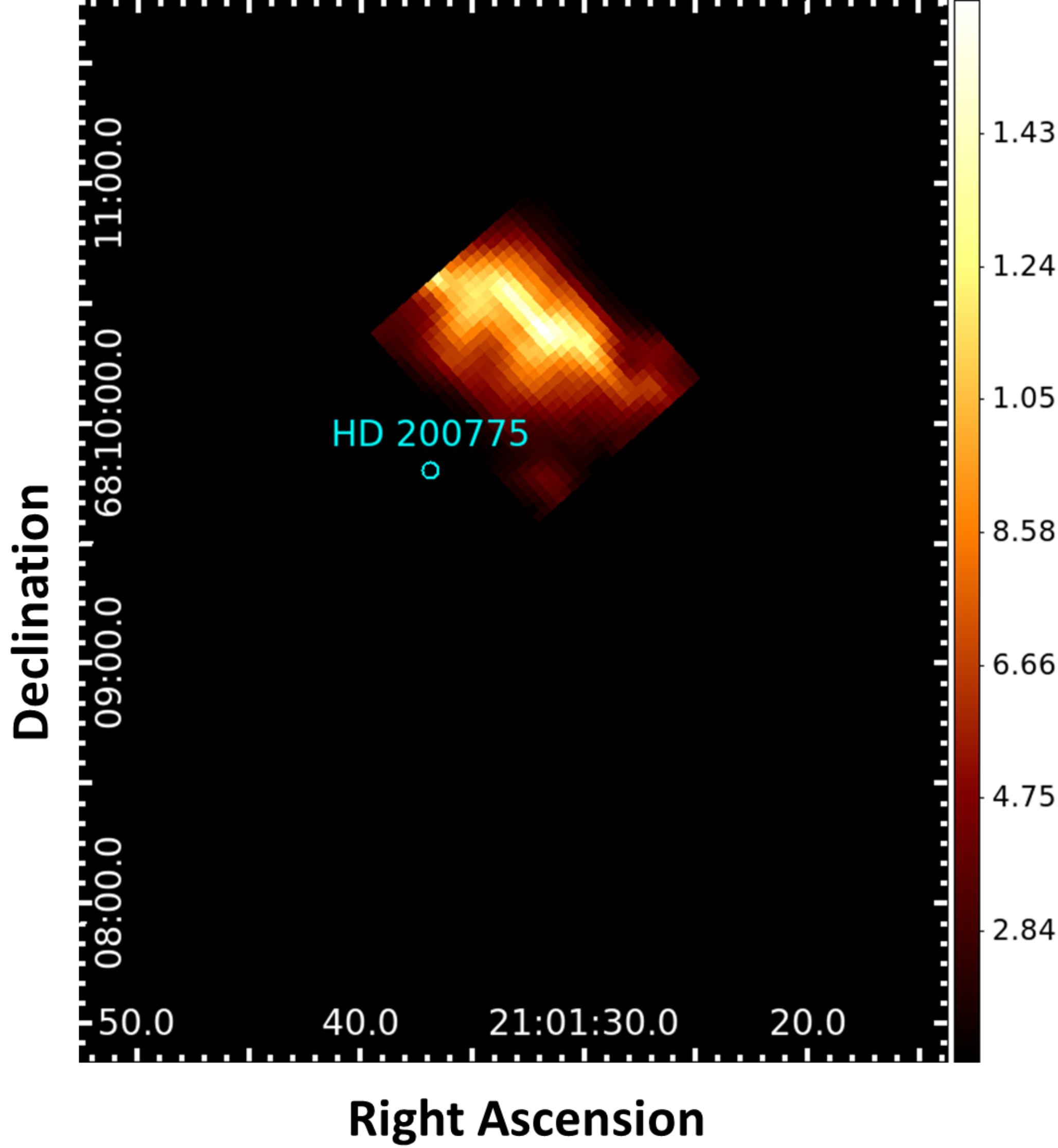}

\end{center}
\caption{ FLITECAM~3.3~$\mu$m (top left), IRAC~8.0~$\mu$m (top right), and the 11.2~$\mu$m PAH band extracted from IRS~SH (bottom center) observations of NGC~7023. The illuminating star, HD~200775, is indicated by a cyan circle, the northwest IRS~SH FOV by a white--dashed rectangle, the ISO-SWS FOV by a black rectangle and the crosscut used is shown as a white line.  We use the same units as given in Figure \ref{ngc2023_images}.}
\label{ngc7023_images}
\end{figure*}

\paragraph{NGC~7023} NGC~7023 is another RN that is well-known for its strong MIR emission \citep[e.g.][]{ces96,boi01,an03,wer04b,rap05,ber12,boe13,boe14a}. It is illuminated by the Herbig Be star, HD~200775, and is located at a distance of 361~$\pm$~6 pc \citep{gaia16,gaia18b}. There are three main regions of study located around this star: 40$^{\prime\prime}$  northwest, 70$^{\prime\prime}$  south, and 170$^{\prime\prime}$  east \citep{ber07}. Each of these regions coincides with a PDR front. In Figure \ref{ngc7023_images}, only the NW and S PDR are visible. The NW PDR is the most prominent of the three with an edge-on structure in which clear stratification of emission is visible \citep[e.g.][]{pil12}. Estimates for the UV radiation field strength at the NW PDR front are 2600  times that of the average interstellar value with gas densities on the order of 4$\times$10$^{3}$ cm$^{-3}$ \citep{cho88}. Due to its prominence, the NW PDR of NGC~7023 is the focus of our study of this RN.

\begin{figure*}[htb]
\begin{center}
\resizebox{\hsize}{!}{%
\includegraphics[clip,trim =.0cm 0.0cm .0cm 0.0cm,width=7.5 cm]{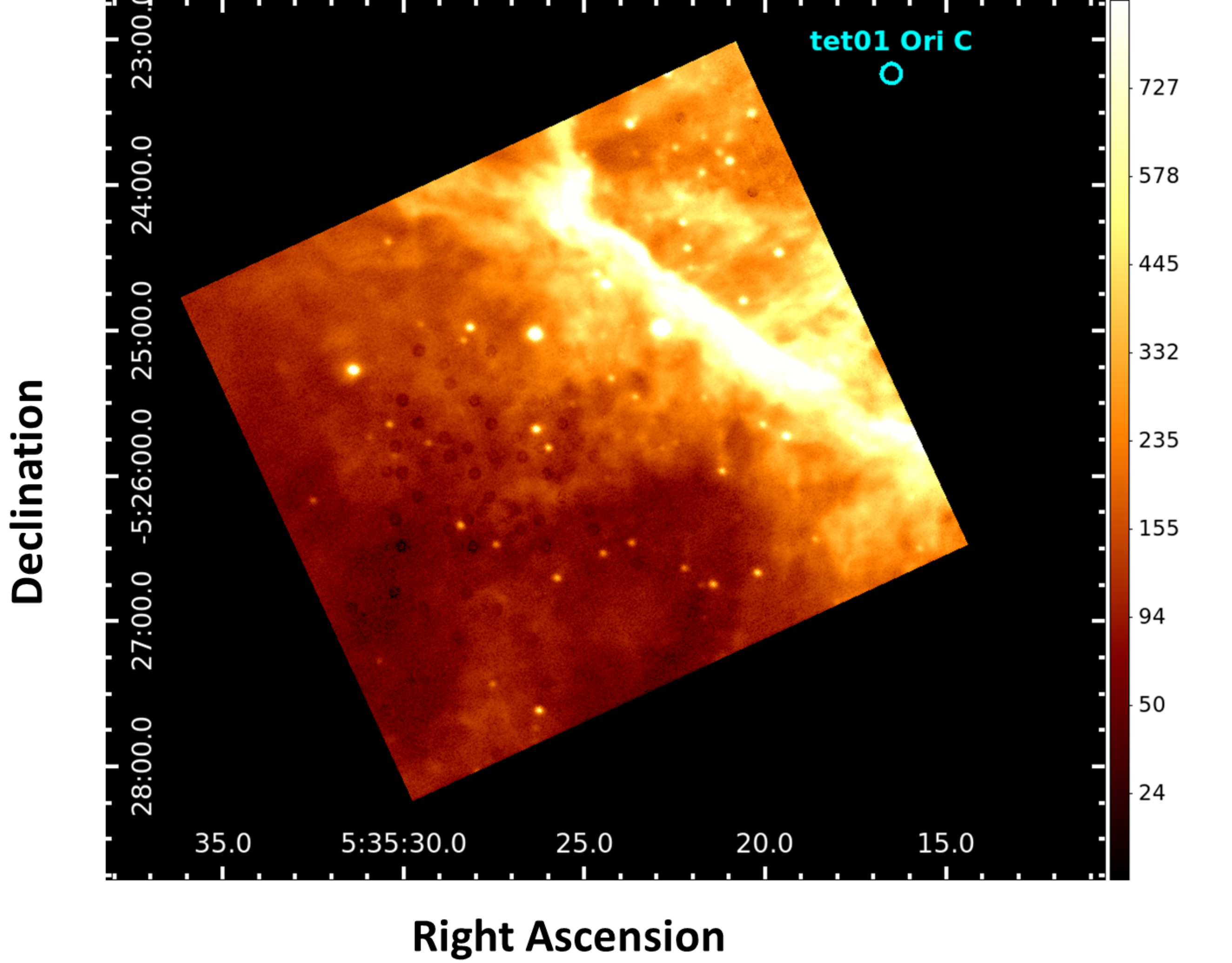}
\includegraphics[clip,trim =.0cm 0.0cm .0cm 0.0cm,width=7.5 cm]{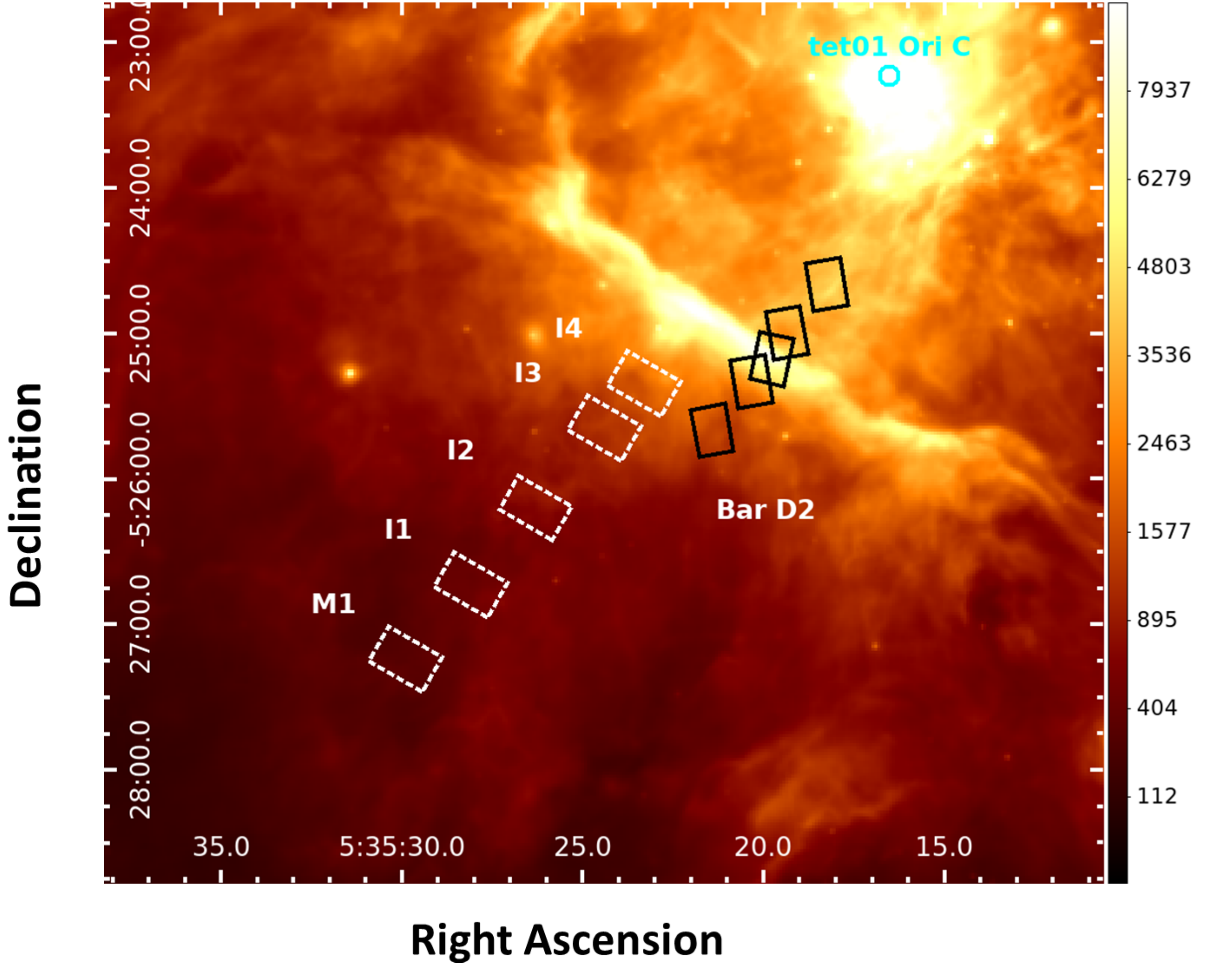}
}
\end{center}
\caption{FLITECAM~3.3~$\mu$m (left) and IRAC~8.0~$\mu$m (right) observations of  the Orion Nebula. The position of the illuminating source, $\theta^{1}$ Ori C, is indicated by a cyan circle, the IRS~SH maps for each position specified in \cite{rub11} by white--dashed rectangles, and the ISO-SWS FOVs by black rectangles. We note that the FLITECAM~3.3~$\mu$m and IRAC~8.0~$\mu$m images use a square root scale  to show the entire range in emission structure.  We use the same units as given in Figure \ref{ngc2023_images}.}
\label{Orion_images}
\end{figure*}

\paragraph{Orion Bar} The Orion Bar or Bright Bar refers to the ionization ridge in the Orion Nebula or the interface between the \HII\ region and PDR located at a distance of 414~$\pm$~7 pc \citep{men07}. The MIR emission within the Orion Bar has been well studied showing a stratified plane-parallel structure in which different species peak at a range of distances from the source  of UV radiation, $\theta^{1}$ Ori C \citep[e.g.][]{ait79,sell81,bre89a,geb89,sell90,tie93,gia94, ces00, rub11,boe12, har12}. In Figure \ref{Orion_images}, the bar structure is quite prominent as it separates the extremely bright Trapezium Cluster responsible for the production of the strong UV radiation field in the northwest from the relatively colder, lower density region extending southeast from the Bright Bar. 
PDR models of the Orion Bar suggest a gas density of 5~$\times$~10$^{4}$ cm$^{-3}$ and FUV radiation field strength of 4~$\times$~10$^{4}$ times that of the average interstellar value \citep[e.g.][]{tau94}. The UV field of the \HII\ region surrounding $\theta^{1}$ Ori is expected to be too harsh to allow a significant PAH population to prosper. Indeed, most of the MIR emission to the north of the Orion Bar is associated with warm dust within the ionized gas \citep{sal16}. Thus our study of the Orion Nebula extends beyond the PDR front within the Bright Bar southwards towards the outer Veil.

\renewcommand{\arraystretch}{1.5}
\begin{table}[tbp]
\caption{\label{table:1} Log of observations.}
\begin{center}
\footnotesize
\begin{tabular}{p{2cm} p{.4cm}  p{1.6cm} p{1.6cm}  p{1.cm} p{.5 cm} }
\hline\hline
Data & S/N$^1$ & RA$^2$ & DEC$^2$  & Nod &  Exp.$^{12}$\\
& & & & ($^{\prime\prime}$ , $^{\circ}$)$^{3}$ & (s)  \\
\hline\\[-15pt]
\multicolumn{5}{c} {NGC~2023}\\[2pt]
FC$^{4}$	&	12 & 5 41 37.89 &  -2 15 52.2  &   1000, 245 & 460 \\
IRS~SH$^{5}$ &  & 5 41 37.63 & -2 16 42.5 &  & \\
IRAC~8.0$^{6,7}$ & 46 & 5 41 37.89 &  -2 15 52.2  & &\\
\hline\\[-15pt]

\multicolumn{5}{c}{NGC~7023}\\[2pt]
FC$^{4,8}$ &  76 & 21 01 36.92 &  68 09 47.7  &  600, 0 & 50  \\
IRS~SH$^{9}$ &  & 21 01 32.89  & 68 09 52.5  & & \\
IRAC~8.0$^{6, 8}$ 	& 80 & 21 01 36.92 &  68 09 47.7 & &\\[2pt]
\hline\\[-15pt]

\multicolumn{5}{c}{ Orion }\\[2pt]
FC$^{4}$ & 40 &  5 35 26.00 &  -5 26 12.0 & 1500, 50 & 360\\
IRS~SH$^{10}$ & & & & & \\
I4 &  & 5 35 23.3 & -5 25 20.7  & & \\
I3 &  & 5 35 24.4 & -5 26 39.1  & & \\
I2 &  & 5 35 26.3 & -5 26 12.1  & & \\
I1 &  & 5 35 28.1 & -5 27 43.8  & & \\
M1 &  & 5 35 29.9 & -5 27 14.4  & & \\
IRAC~8.0$^{11}$ & 39 &  5 35 22.33 &  -5 24 36.0 & &\\[2pt]
\hline\\[-20pt]

\end{tabular}

\end{center}
$^1$ This value corresponds to the peak SNR in each frame.;
$^2$ $\alpha$, $\delta$ (J2000) refer to the center of the map; $\alpha$ has units of hours, minutes, and seconds, and  $\delta$ has units of degrees, arc minutes, and arc seconds;
$^3$ Nod parameters (throw, angle);
$^4$  FLITECAM;
$^5$ e.g. \cite{pee12,pee17,sha16};
$^6$ Spitzer Heritage Archive;
$^7$ \cite{fle10};
$^8$ \cite{cro16};
$^{9}$ e.g. \cite{ber07,boe13,sto16};
$^{10}$ \cite{rub11,boe12};
$^{11}$ \cite{meg12}; $\alpha$, $\delta$ in this case reference the center of a sub image extracted from the original mosaic.
$^{12}$ Exposure Time. 
\vspace{0.5cm}
\end{table}

\section{Observations}
\label{obs}
We used multiple MIR observations to measure the dominant PAH emission bands. First, we obtained photometric observations from the First Light Infrared TEst CAMera \citep[FLITECAM,][]{tem12} on--board the Stratospheric Observatory for Infrared Astronomy \citep[SOFIA,][]{you12} to probe the 3.3~$\mu$m PAH emission. Second, PAH emission in the 7--9~$\mu$m range was investigated using observations from the Infrared Array Camera \citep[IRAC,][]{faz04} on--board the Spitzer Space Telescope \citep{wer04a}. 
Finally, the 11.2~$\mu$m PAH emission was extracted from observations in the short-high mode (SH) of the Infrared Spectrograph \citep[IRS,][]{hou04} on--board the Spitzer Space Telescope. 
These observations are detailed below and are summarized in Table \ref{table:1}.

\subsection{SOFIA observations}
\label{SOFIA}

We obtained SOFIA-FLITECAM observations in the 4$^{th}$ to 6$^{th}$ cycles for both NGC~2023 (Figure \ref{ngc2023_images}) and towards the southeast of the Orion Bar (Figure \ref{Orion_images}) in the `PAH' filter, which has an effective wavelength of 3.302~$\mu$m and a bandwidth of 0.115~$\mu$m (PID: 04\_0058, PI: A. Tielens). The FLITECAM instrument has a 1024~pixel~$\times$~1024~pixel InSb detector that covers a 8$^\prime$~$\times$~8$^\prime$ area on the sky with 0.475$^{\prime\prime}$ ~$\times$~0.475$^{\prime\prime}$~pixels.

The NGC~2023 field of view (FOV) is centered on the central star, HD~37903, while the Orion FOV was chosen such that it is centered on the Spitzer-IRS~SH pointing I2 as defined in \cite{rub11}, which probes the region southeast of the Orion Bar, in order to cover the five Spitzer--IRS~SH pointings closest to the Trapezium cluster observed by \cite{rub11}.  
To account for background emission, the instrument was used in nod mode, where the nod parameters were set so that the telescope was moved sufficiently off source between each observation. In the case of NGC~7023 (Figure \ref{ngc7023_images}), we have used SOFIA-FLITECAM archival data previously published by \cite{cro16}. The FLITECAM~3.3~$\mu$m images were found to have a point source FWHM of $\sim$~2.1$^{\prime\prime}$ .
 
\subsection{Spitzer observations}
\label{Spitzer}

We obtained Spitzer-IRAC~8.0~$\mu$m observations (channel 4) of each source (see Figures \ref{ngc2023_images}, \ref{ngc7023_images}, and \ref{Orion_images}). The IRAC~8.0~$\mu$m band is a broadband filter with a bandwidth of 2.9~$\mu$m and a nominal wavelength of 7.87~\mum, and is found to be an excellent tracer for the $\Sigma$7--9~$\mu$m PAH emission \citep[e.g.][ see Appendix \ref{appa}]{smi07, sto14}. The IRAC~8.0~$\mu$m observations were found to have a point source FWHM of 2$^{\prime\prime}$.

For each source, we measured the 11.2~$\mu$m PAH emission using IRS~SH data, which covers the wavelength range of 10--20~$\mu$m at a  spectral resolution of 600. IRS~SH spectral maps were obtained over a portion of the nebula for both NGC~2023 and NGC~7023 (see Figures \ref{ngc2023_images} and \ref{ngc7023_images} respectively). The spectral map of NGC~2023 covers the S and SSE emission ridges, which correspond to the PDR front as traced by H$_{2}$ emission (following the nomenclature of \citealt{pee17}), as well as YSO C defined by \citet{sell83}. The spectral map of NGC~7023 is comprised of the region surrounding the NW PDR. For Orion, we use IRS~SH pointings taken along a line from 2.1 to 5.1$^\prime$  extending radially away from the illuminating source $\theta^{1}$ Ori C near the Bright Bar \citep{rub11,boe12}. The spectra obtained for each pointing correspond to a 2~$\times$~10  aperture grid pattern or `chex' of SH spectra \cite[for details, see][]{rub11}. The resulting FOV of each of these pointings (25.4$^{\prime\prime}$~$\times $~16.3$^{\prime\prime}$) and the position of each pointing are shown in Figure \ref{Orion_images}. These pointings are given the following designations by \cite{rub11} with increasing distance from the  Bar: `inner' (I4, I3, I2, I1), and `middle' (M1).

\section{Data Processing}
\label{data}

The same general reduction procedure was used for all sources to generate emission ratio maps. 

\subsection{FLITECAM}

To extract the 3.3~$\mu$m emission feature, the raw FLITECAM frames were co-added using a noise weighted mean into a single image and flux calibrated using standard stars observed during the same night as the observations. These images were slightly misaligned with respect to the IRAC~8.0~$\mu$m images. We correct for this misalignment by shifting our FLITECAM images to align with the IRAC~8.0~$\mu$m images, this is done by aligning the prominent stellar sources or the extended emission peaks in the FLITECAM frames with the corresponding sources in the IRAC~8.0~$\mu$m frames. The FLITECAM images are converted from units of surface brightness (Jy/pixel) to flux (W~m$^{-2}$~sr$^{-1}$) by multiplying by the bandwidth of the FLITECAM filter, assuming a nominal flat spectrum. This is done in order to have comparable units with the 11.2~$\mu$m emission extracted from the IRS~SH spectra.

\subsubsection{Reflection Nebulae}
\label{RNe_FC}
In the case of NGC~7023, we obtained a single set of observations for which it was necessary to rotate the FLITECAM image to get the NW and S PDRs to overlap with respect to the IRAC image  \citep[][]{cro16}. 
In the case of NGC~2023, we obtained three sets of observations. Each of these three FLITECAM frames had a significantly high median on-frame background level that is subtracted from each frame (755, 357, and 224~MJy/sr) after which they are combined in a single image.
We checked the absolute calibration for both sources in two ways. First, by comparison with ISO-SWS observations\footnote{The TDT (Target Dedicated Time) number is 20700801 and 65602309 for respectively NGC~7023 and NGC~2023.}. In particular, we  checked the absolute flux calibration of the ISO-SWS spectra with IRAC~3.6~$\mu$m observations by comparing the observed IRAC~3.6~$\mu$m emission with the expected value in the IRAC~3.6~$\mu$m bandpass based on the ISO-SWS spectrum. We found a consistency of 90\% and 98\% for NGC~7023 and NGC~2023 respectively, indicating that the ISO-SWS spectra are well calibrated in an absolute sense. Subsequently, we compare the FLITECAM emission observed in the ISO-SWS aperture with the expected FLITECAM emission based on the ISO-SWS spectrum. We found an agreement of 100\% and 104\% for NGC~7023 and NGC~2023 respectively, indicating that the FLITECAM images are well calibrated. Second, by comparison with photometry of their central star. Specifically, for NGC~2023, we performed photometry on the central star, HD~37903, in each frame after background subtraction which resulted in a flux density of 0.38~$\pm$~0.02 Jy on average. Similarly for NGC~7023, we performed photometry on the central star, HD~200775, and found a flux density of 11.56~$\pm$~0.03~Jy. This is consistent with the spectra of HD~37903 and HD~200775 given in \cite{sell84}, again demonstrating that our FLITECAM data is properly calibrated.

We estimate the PAH contribution to the total flux observed in this FLITECAM filter from AKARI spectral observations of NGC~7023 coinciding with the NW PDR \citep{pil15}.  We do this by multiplying both the continuum--subtracted 3.3~$\mu$m band and the total spectrum with the FLITECAM~3.3~$\mu$m filter response curve and subsequently integrating over the filter range for each AKARI pointing. Dividing these quantities, we found that the PAH contribution in the 3.3~$\mu$m filter is 59--74\% of the total emission (which comprises both PAH and dust continuum emission). As both continuum and PAH features are due to UV pumped fluorescence from large molecules \citep{all89} and both species require similar sizes to be excited, it is reasonable to assume that the feature to continuum ratio will remain constant within a given source. Moreover, ISO-SWS observations of NGC~2023 and NGC~7023 show that both sources have very similar feature to continuum emission ratios for the 3.3~$\mu$m band \citep[e.g.][their figure 3]{mou99}, thus the total PAH contribution in the FLITECAM filter will be approximately the same.

\subsubsection{Orion}
\label{Orion_FC}

\begin{figure}[h!]
\begin{center}
\includegraphics[clip,trim =1cm 0.5cm 0cm 2cm,width=8.8cm]{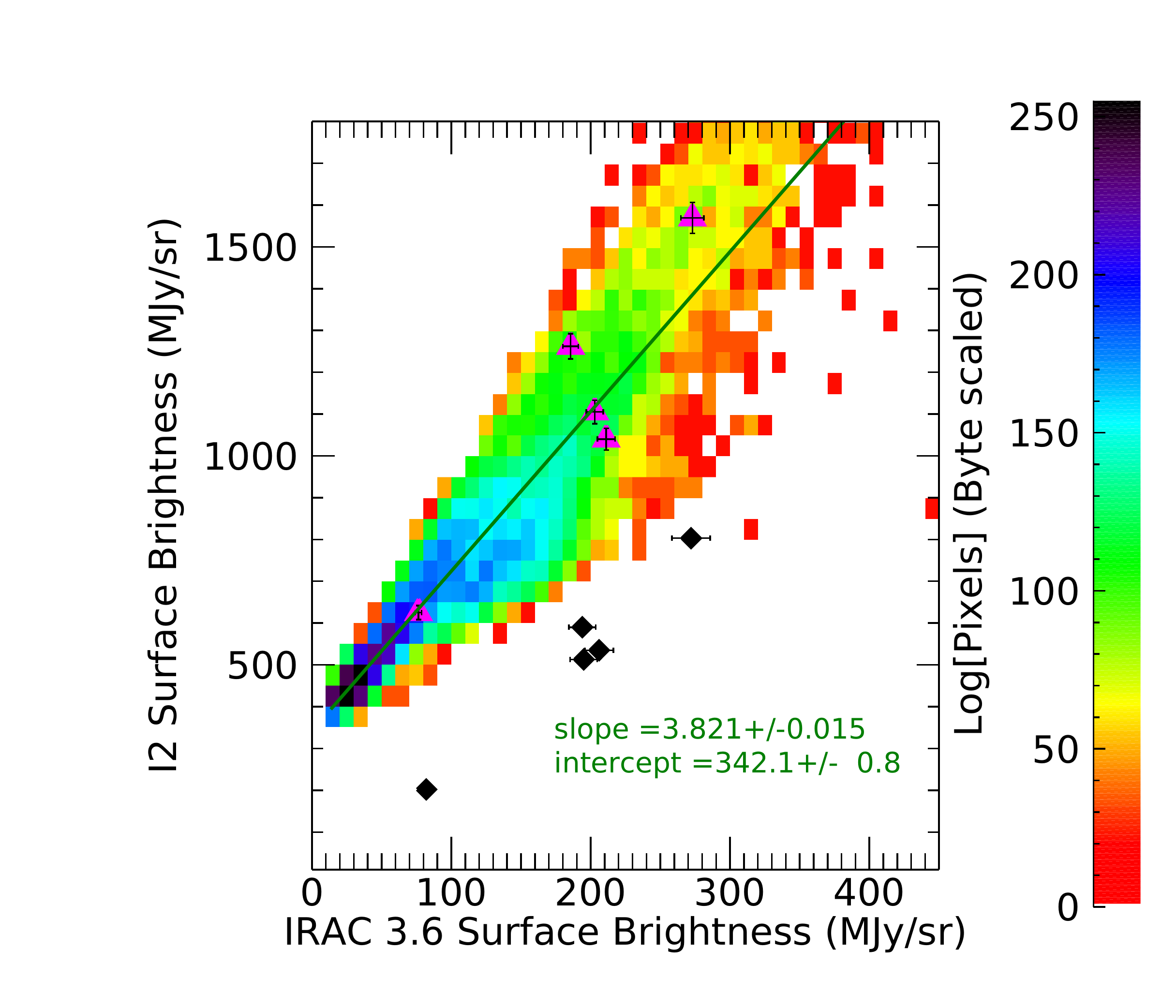}
\end{center}
\caption{ The observed surface brightness in the FLITECAM data and the IRAC~3.6~$\mu$m data (after extended source correction). All stellar sources within the FOV have been masked. Surface brightness values of the FLITECAM observations within the ISO-SWS apertures are indicated by magenta triangles and expected mean surface brightness values (from multiplying the SWS spectra with each filter response curve) are shown as black diamonds. A weighted linear fit is given by green line.}
\label{orion_fc_fit}
\vspace{-0.2cm}
\end{figure}

We investigate the absolute calibration of the Orion FLITECAM~3.3~$\mu$m data by comparing it with IRAC~3.6~$\mu$m data. First, we estimate the expected ratio of the FLITECAM to IRAC observations for the Orion Bar. To this end, we use the ISO-SWS spectra across the Orion Bar\footnote{There are five ISO-SWS spectra available across the Orion Bar at distances of 2.597$^\prime$, 2.175$^\prime$, 1.971$^\prime$, 1.766$^\prime$, and 1.356$^\prime$ from $\theta^{1}$ Ori C referred to as positions D2, H2S1, D5, Br$\gamma$, and D8 respectively \citep[Figure~\ref{Orion_images}, e.g.][]{pee02, vdie04}. Their TDT numbers are 69502005, 69501806, 83101507, 69502108, and 69501409 respectively.}.
We multiply these ISO-SWS spectra with both filter response curves and find that the average FLITECAM-to-IRAC ratio equals 2.7~$\pm$~0.3. Next, we compare the observed surface brightness of the FLITECAM 3.3~$\mu$m and the IRAC~3.6~$\mu$m images (after applying the IRAC~3.6~$\mu$m extended source correction; Figure \ref{orion_fc_fit}). The weighted line of best fit to the data has a slope of 3.82~$\pm$~0.02 and a y--intercept of 342.1~$\pm$~0.8~MJy/sr. We thus find a scaling factor of 1.4~$\pm$~0.2 which is equal to the observed FLITECAM-to-IRAC ratio divided by the expected FLITECAM-to-IRAC ratio.   Hence, to calibrate the FC (FLITECAM~3.3~$\mu$m) observations, we use the following scaling relation:

\begin{equation}
FC(\textrm{scaled)} = (FC - 342.1)/1.4,    
\end{equation}

Applying this relationship to the observed Orion FLITECAM~3.3~$\mu$m data yields values that agree with the expected FLITECAM~3.3~$\mu$m values based on the SWS spectra within 12\%.

To further validate this approach, we also apply it to the RNe. The predicted FLITECAM to IRAC ratio is determined by multiplying the AKARI spectra of NGC~7023 \citep{pil15} with the respective filters. We find that the mean flux density of the spectrum multiplied with the 3.3~$\mu$m filter is about 2 times greater than multiplying with the IRAC~3.6~$\mu$m. A similar ratio is expected for NGC~2023 based on the similarity of the IR spectra of these sources \citep[e.g.][]{mou99,ver01}. Using the same sized apertures as the Orion pointings in both RNe, we are able to reproduce the predicted FLITECAM-to-IRAC flux ratios of 2. This shows that the FLITECAM observations of the RNe are in agreement with predicted values.

We estimate the PAH contribution to the total flux observed in the FLITECAM filter from the ISO-SWS Orion Bar D2 position by multiplying the filter response curve with the Bar D2 spectrum \citep[e.g.][]{pee02, vdie04}. We find that the total PAH contribution to the 3.3~$\mu$m filter is about 71\% of the total emission (which comprises both PAH and dust continuum emission).  

\newpage

\subsection{Spitzer}
\label{spit}
We obtained the Spitzer-IRS data in reduced form (for details on the reduction process, see \cite{pee12} and \cite{sha15} for the RNe and \cite{boe12} for the Orion data). To measure the 11.2~$\mu$m PAH emission, we defined and subtracted a local spline continuum with anchor points at 10.85 and 11.65~$\mu$m \cite[as in][]{pee17} from the spectrum at each pixel in our spectral maps or in the case of the Orion IRS~SH data, from the spectrum at each pointing. We performed a simultaneous Gaussian fitting procedure to the 11.0 and 11.2~$\mu$m emission features in each continuum--subtracted spectra in order to separate these features \citep[e.g.][]{sto14,sto16,pee17}. The flux of the 11.2~$\mu$m emission band is then determined by integrating these continuum--subtracted spectra over the wavelength range given above and subtracting the 11.0~$\mu$m band flux.\\

We obtained the Spitzer-IRAC data in reduced form for NGC~2023 and the Orion Nebula \citep{fle10, meg12} and retrieved them from the Spitzer Heritage Archive in the case of NGC~7023. 
To approximate the 7.7~$\mu$m PAH emission, we converted the IRAC~8.0~$\mu$m images (MJy~sr$^{-1}$) to flux densities (W~m$^{-2}$~sr$^{-1}$) by multiplying by the wavelength range covered by the 7.7~$\mu$m emission band of $\sim$~1.0~$\mu$m \citep[e.g.][]{pee02}, again assuming a nominal flat spectrum. The relative PAH contribution in this filter is about 89\% of the total emission in the diffuse ISM \citep{sto14} and between 72--80\% across the S ridge in NGC~2023 over the distance range for which we have 11.2~$\mu$m IRS~SH measurements (Appendix \ref{appa}). \\

In order to compare the spatial distribution of the different PAH bands for the RNe, the FLITECAM and IRAC~8.0~$\mu$m images were regridded to the IRS~SH pixel scale (2.3$^{\prime\prime}$~$\times$~2.3$^{\prime\prime}$ ). Both images were additionally averaged over 2~$\times$~2~pixels, set by the IRS spatial resolution. Bright stellar sources in the IRS FOVs and pixels for which there was no 3~$\sigma$ detection of the 11.2/3.3 PAH ratio were masked out. In case of the Orion, we determined the median flux density for both FLITECAM and IRAC~8.0~$\mu$m images within the IRS aperture at each pointing. In this way, the PAH emission at 3.3, 7.7, and 11.2~$\mu$m are compared at 5 different pointings within the Orion region.

\begin{figure*}[th!]
\begin{center}
\includegraphics[clip,trim =1cm 0.5cm .5cm 0cm,width=7.cm, angle =33.16 ]{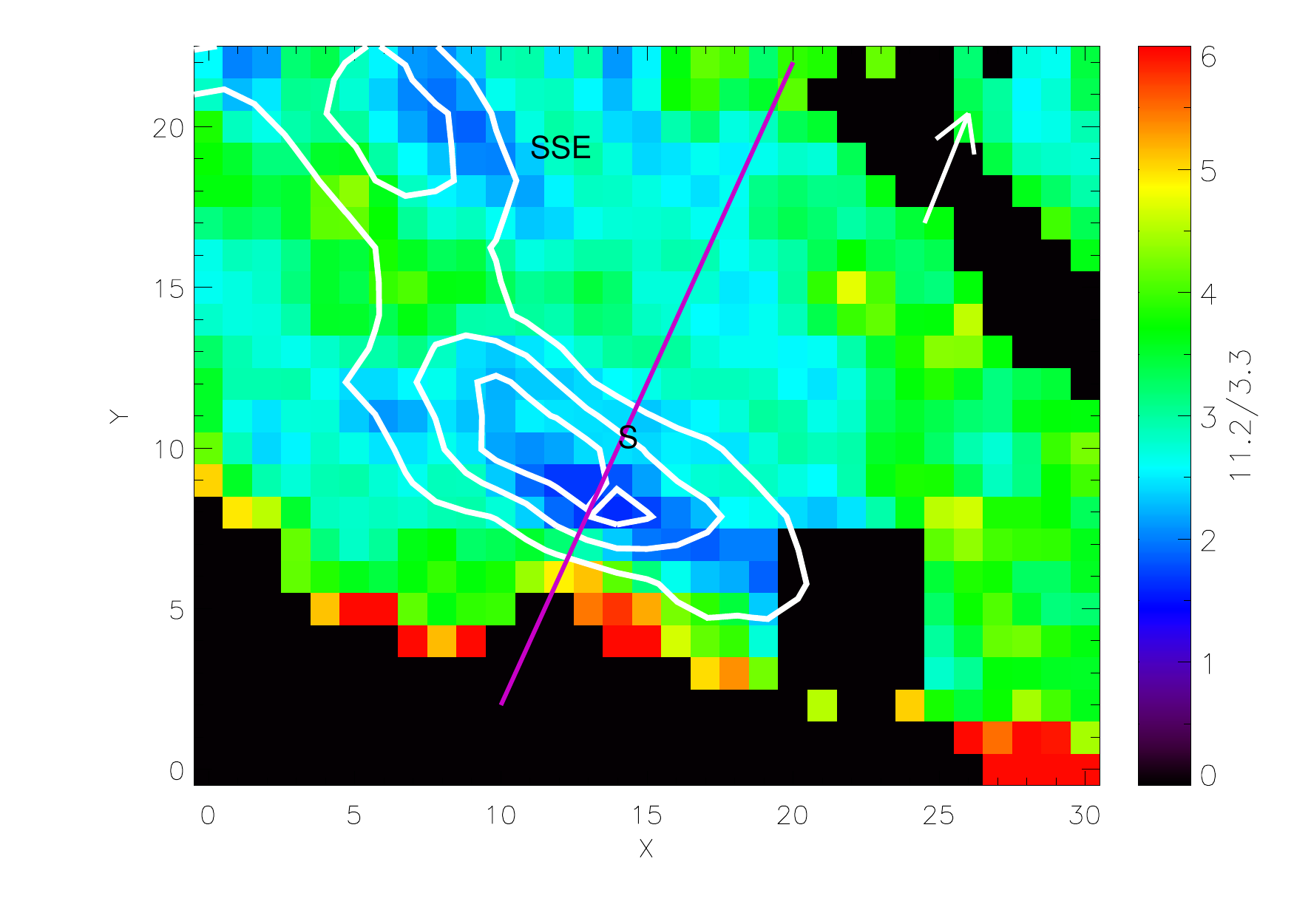}
\includegraphics[clip,trim =1cm 0.5cm .0cm 0cm,width=7.cm,  angle =33.16]{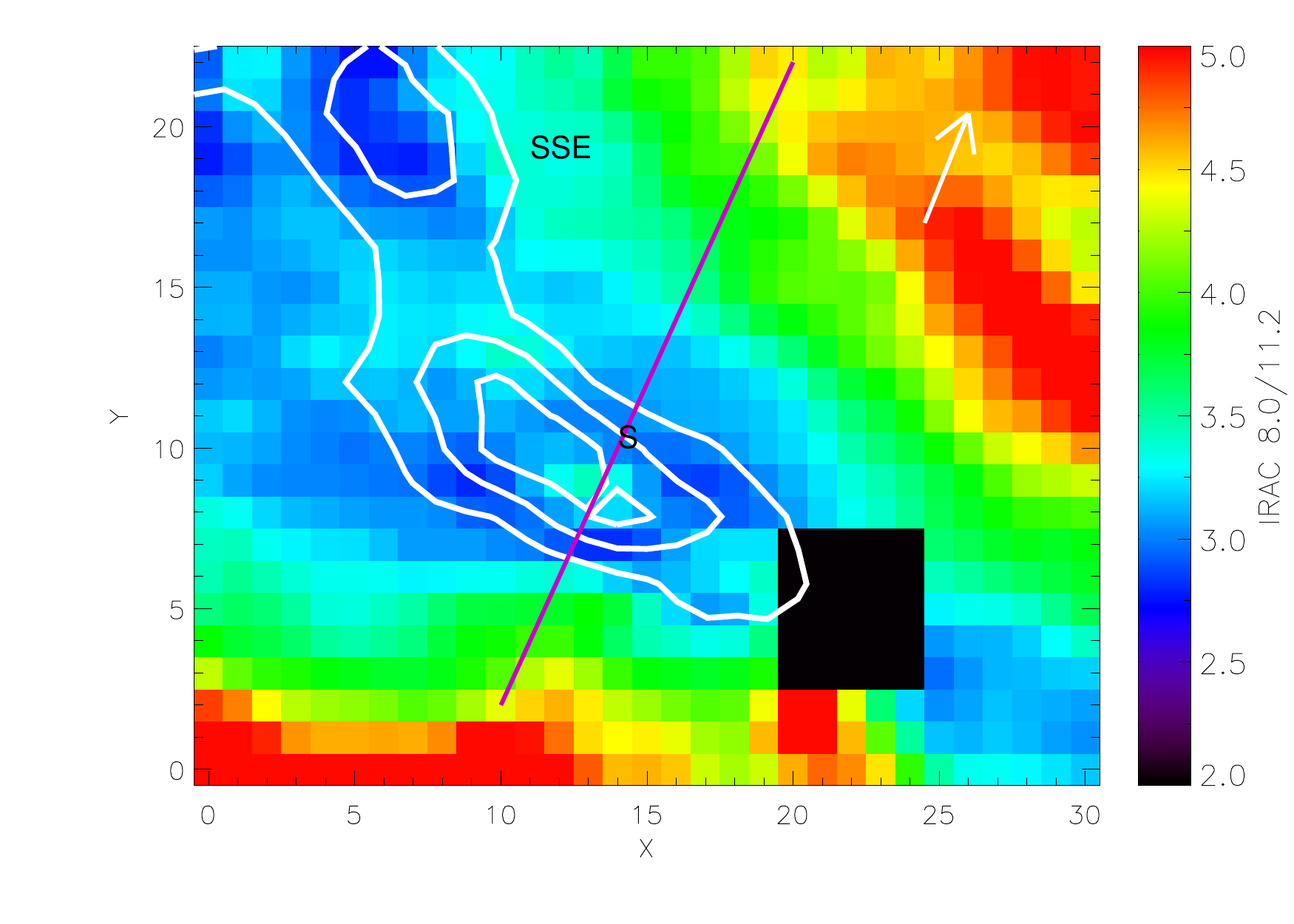}

\end{center}
\caption{The 11.2/3.3 (left panel) and the IRAC~8.0/11.2 (right panel) PAH intensity ratios for the south FOV in NGC~2023. Contours of the 12.3~$\mu$m 0-0 S(2) H$_{2}$ line are overplotted in white ((2.5, 3.32, 4.14)~$\times$~10$^{-7}$~W~m$^{-2}$~sr$^{-1}$). The direction to the illuminating source, HD~37903, is indicated by a white arrow. A magenta line is overplotted to indicate the position of the radial profile used for this source. The MIR emission peaks referred to as the the south ridge and south-southeast ridge as in \citet{pee17} are given by S and SSE respectively. Pixels below a 3~$\sigma$ detection or near the source YSO C are set to zero (shown here in black). North is up and east is to the left.}
\label{ngc2023_ratios}
\end{figure*}

\begin{figure*}[th!]
\begin{center}
\includegraphics[clip,trim =1cm 0.5cm 0.5cm 0cm,width=8.5cm ]{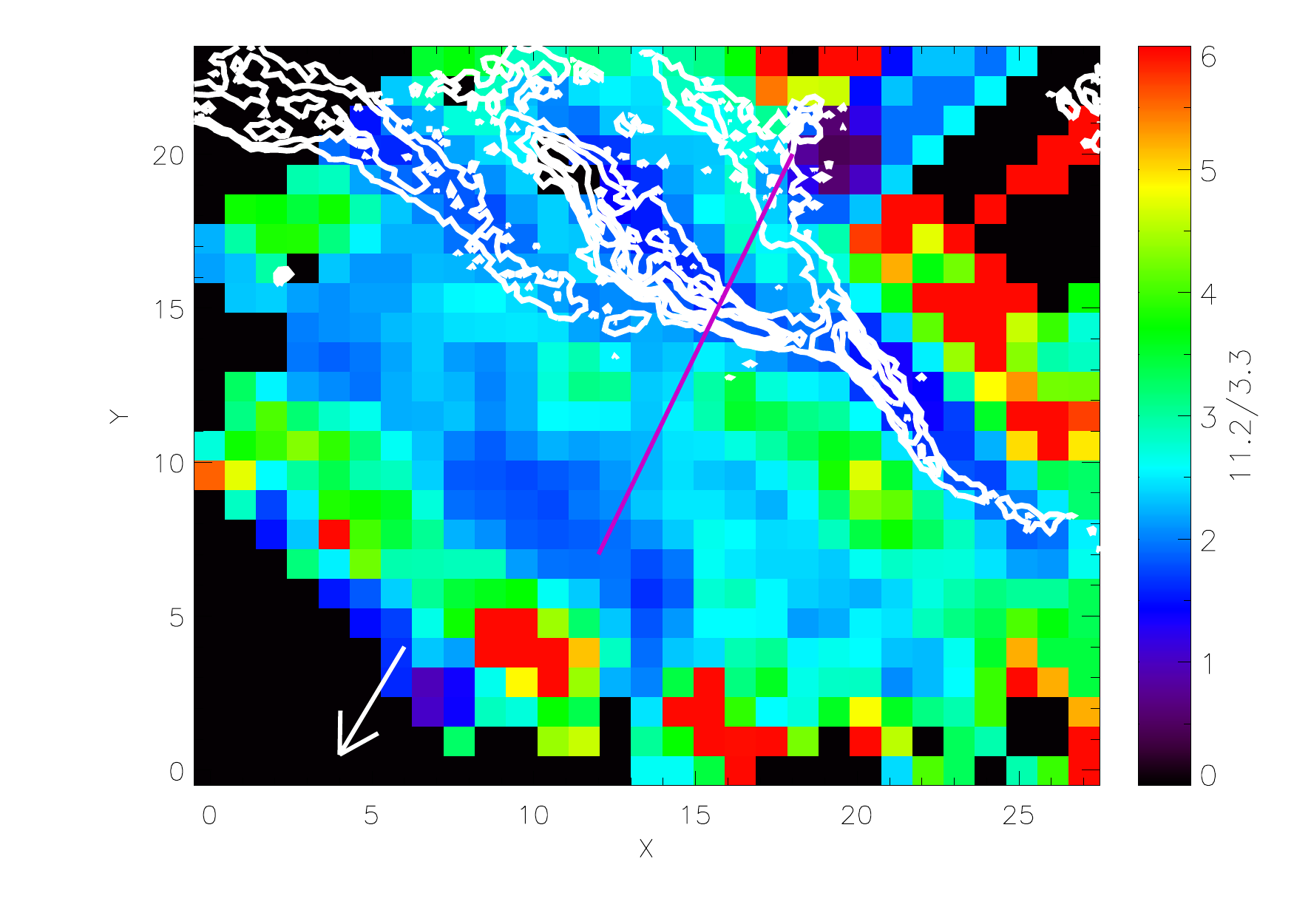}
\includegraphics[clip,trim =1cm 0.5cm 0.5cm 0cm,width=8.5cm]{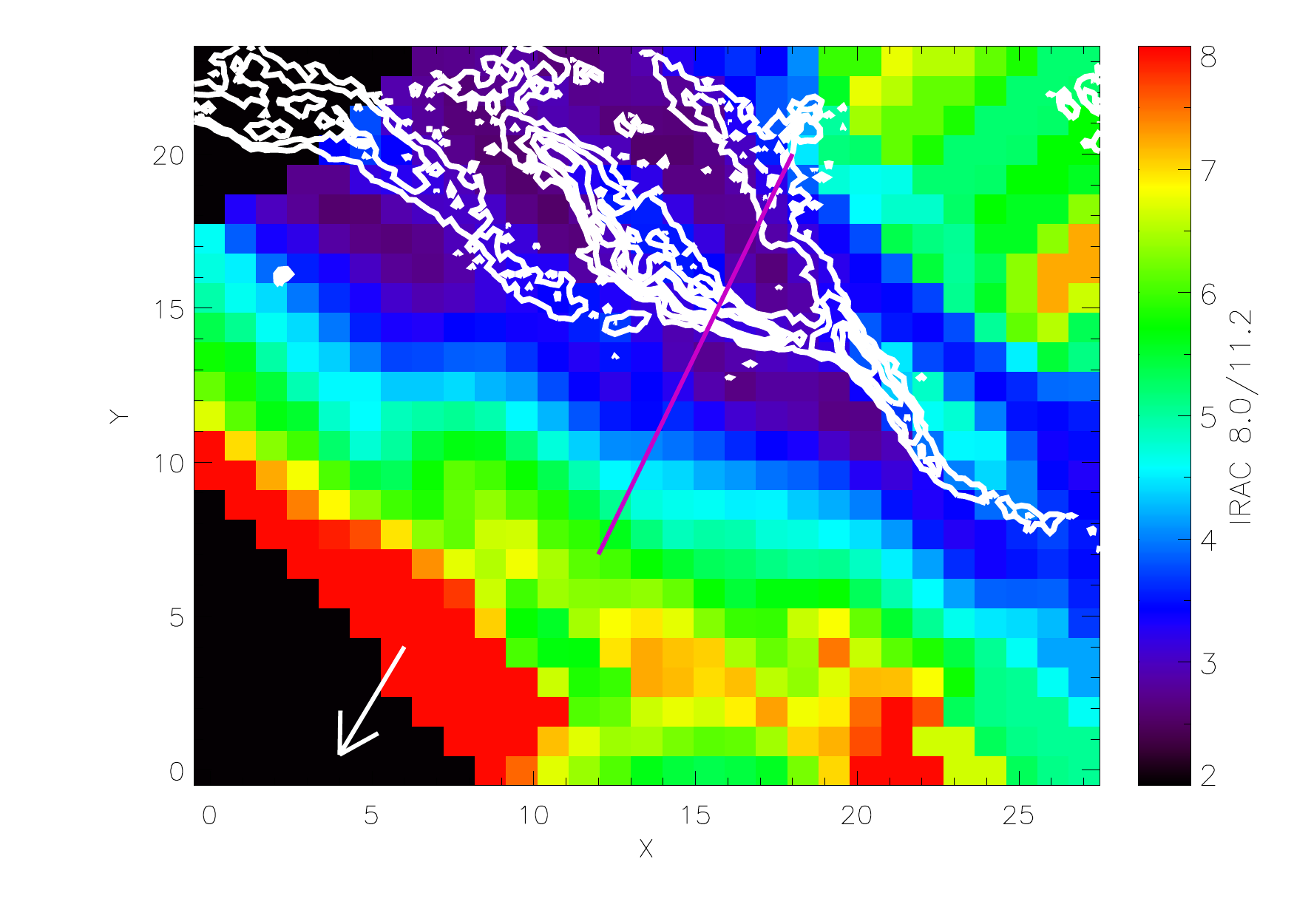}
\end{center}
\caption{ The 11.2/3.3 (left panel) and the IRAC~8.0/11.2 (right panel) PAH intensity ratios for the northwest NGC~7023 FOV. Contours of the 2.124~$\mu$m 1-0 S(1) H$_{2}$ line are overplotted in white \citep[80, 120, 175, 220~arbitrary units,][]{lem96}. The direction to the illuminating source, HD~200775, is indicated by a white arrow. A magenta line is overplotted to indicate the position of the radial profile used for this source. Pixels below a 3~$\sigma$ detection or outside of the IRS~SH aperture are set to zero  (shown here in black).  North is up and east is to the left.}
\label{ngc7023_ratios}
\end{figure*}

\section{Results}
\label{results}

We determined the 11.2/3.3 and IRAC~8.0/11.2 (as probed by IRAC~8 $\mu$m/11.2~$\mu$m) PAH intensity ratios spatially where images of each band were present. For NGC~2023 and NGC~7023, these ratios were measured where each of our observations overlap, this resulted in a FOV of the same size as the IRS~SL maps. For Orion, we determined these ratios in the IRS~SH apertures in covered by the FLITECAM frame. We consider the ratio maps and cuts obtained for each source individually. 

\begin{center}
 {\it NGC~2023}    
\end{center}

Figure~\ref{ngc2023_ratios} shows the 11.2/3.3 and IRAC~8.0/11.2 PAH intensity ratios in the southern PDR of NGC~2023. Contours of the 12.3 $\mu$m H$_{2}$ line are overplotted in white to emphasize the S and SSE emission ridges. These bright filaments are thought to be edge-on corrugations of the cloud material that surrounds the bowl-shaped cavity \citep{fie94}. The central star is beyond the upper right edge of this map. 

 We find minima in the 11.2/3.3 PAH intensity ratios at the peak of the H$_{2}$ emission and the 3.3~$\mu$m emission near the S ridge. In the SSE ridge, the minimum is slightly displaced (in the direction towards the star) compared to the H$_2$ peak intensity, again coinciding with the nearby peak of the 3.3~$\mu$m emission. Another minimum in the 11.2/3.3 ratio behind the H$_2$ peak centered at pixel (5, 11) is also apparent. In general, we find an increase in this ratio as we move towards the star and further into the molecular cloud behind both ridges. Moving inwards towards the star from the S ridge H$_{2}$ peak, there is a local minimum shown in cyan surrounded by green which extends westward from the SSE ridge towards the far right of the image and which is centered at pixel coordinates (15, 18). This ratio then proceeds to rise moving further inwards towards the top right of the frame. 

Overall, the IRAC~8.0/11.2 PAH intensity ratio exhibits a stratified morphology with its highest values closest to the central star. In addition, it exhibits a narrow local peak just above the H$_{2}$ peak within the S ridge as well as in the lower right portion of the S ridge. The SSE ridge, in contrast, has a minimum in this ratio.  Below the S ridge into the molecular cloud, there is a rise in this ratio.

\begin{figure*}[!htbp]
\begin{center}
\resizebox{\hsize}{!}{%
\includegraphics[clip,trim =.8cm .5cm 0.5cm .5cm,width=4cm]{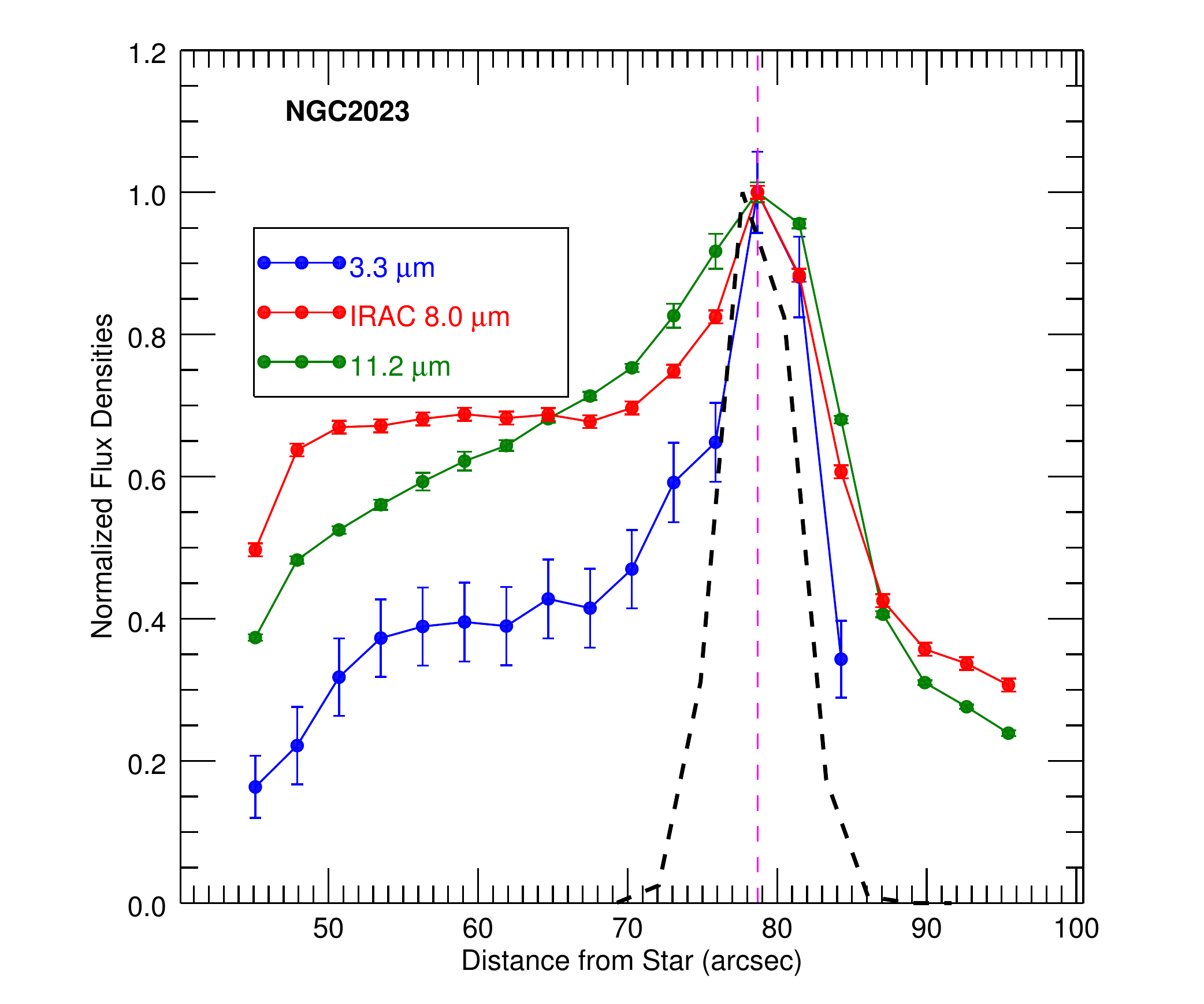}
\includegraphics[clip,trim =.8cm .5cm 0.5cm .5cm,width=4cm]{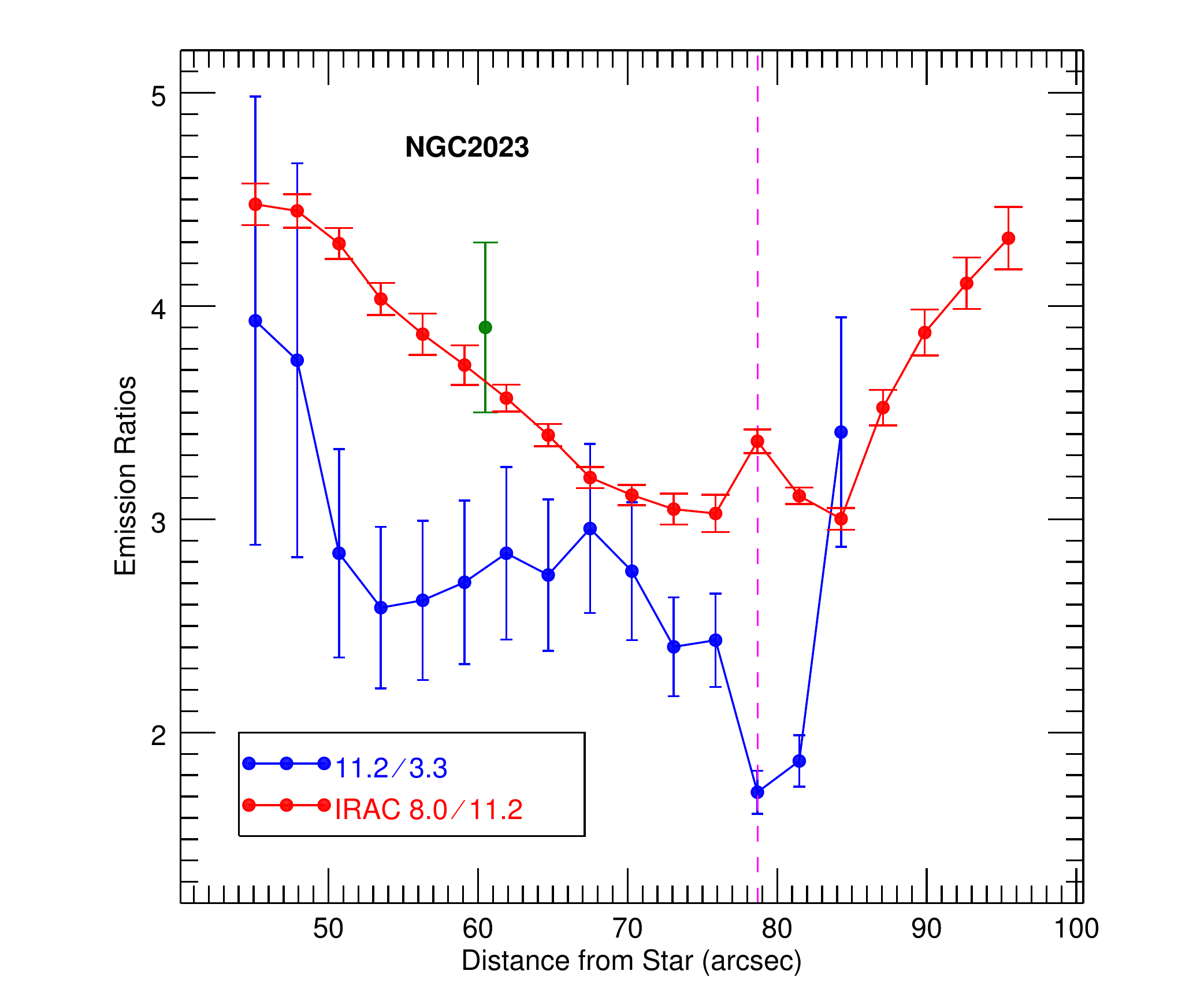}}
\resizebox{\hsize}{!}{%
\includegraphics[clip,trim =.8cm .5cm 0.5cm .5cm,width=4cm]{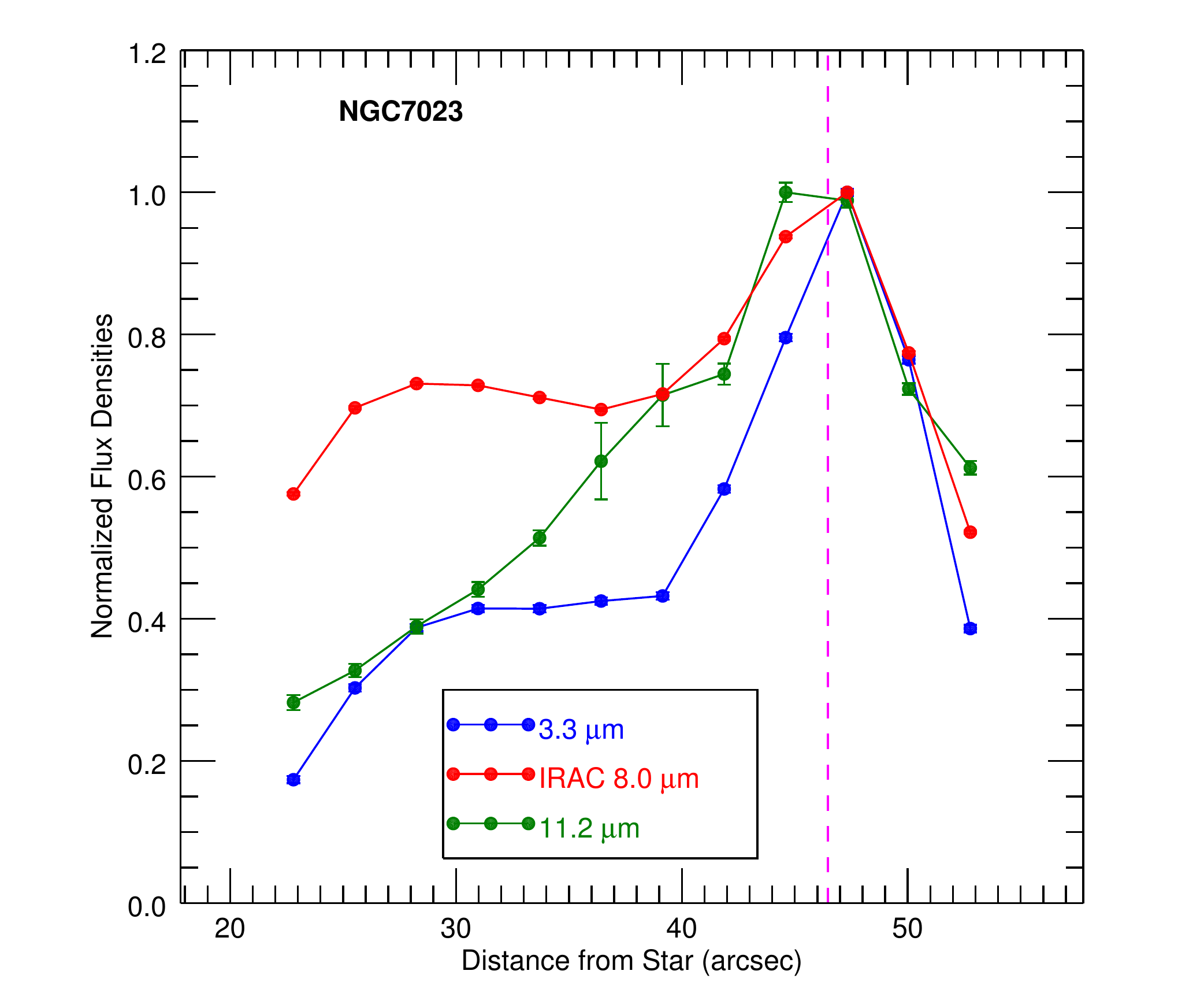}
\includegraphics[clip,trim =.8cm .5cm 0.5cm .5cm,width=4cm]{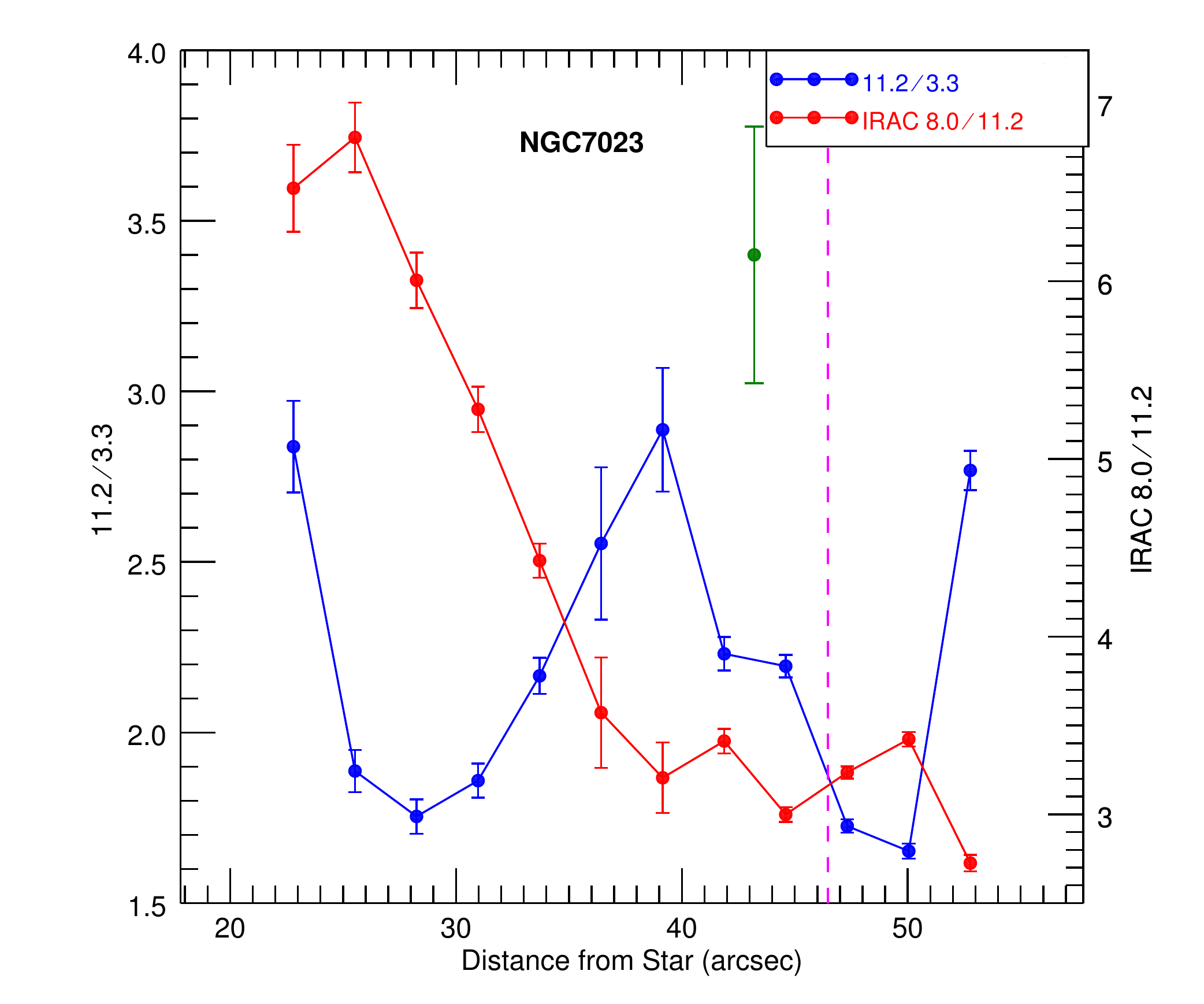}
}
\end{center}
\caption{Shown are normalized radial profiles of 3.3, IRAC~8.0, and 11.2~$\mu$m intensities (left panels), radial profiles of the 11.2/3.3 and the IRAC~8.0/11.2 intensity ratios (right panels) as a function of distance to the central star in NGC~2023 (top, see Figure~\ref{ngc2023_images}) and NGC~7023 (bottom, see Figure~\ref{ngc7023_images}). In the top left panel, we overplot the SH PSF as a black--dashed line. The 11.2/3.3 intensity ratio determined from the ISO-SWS pointing of each source is shown as a green circle in the respective ratio profiles (right panels, see text for details). The PDR front is traced by the 12.3 $\mu$m 0-0 S(2) H$_{2}$ emission line in NGC~2023 and the 2.124 $\mu$m 1-0 S(1) H$_{2}$ emission line in NGC~7023 (shown in magenta). Note that increasing 11.2/3.3 corresponds to increasing PAH size and increasing IRAC~8.0/11.2 corresponds to increasing ionization. The associated uncertainties are indicated as error bars of the same color as the respective data.}
\label{RNe_lineprofiles}
\end{figure*}

Figure~\ref{RNe_lineprofiles} shows a crosscut taken across the S ridge emission peak (see Figure~\ref{ngc2023_images} for the orientation of the cut) of the normalized flux densities of the 3.3, IRAC~8.0, and 11.2~$\mu$m emission as well as the 11.2/3.3 and IRAC~8.0/11.2 ratios. Note that the 3.3~$\mu$m and the 11.2/3.3 profile have a shorter range beyond the PDR front, due to the lack of a 3~$\sigma$ detection with FLITECAM. All three emission features peak at the PDR front (as traced by the peak intensity of the H$_2$ emission, at 78.6$^{\prime\prime}$  from the illuminating source). However, the 3.3~$\mu$m emission drops fastest in intensity away from this PDR front, followed by the 11.2~$\mu$m emission. Interestingly, the IRAC~8.0~$\mu$m emission shows a plateau from $\sim$ 50--70$^{\prime\prime}$ , before dropping off as well closer to the central star. This overall radial profile is reflected in the 11.2/3.3 and IRAC~8.0/11.2 PAH intensity ratios which show, in general, a smooth decrease with increasing distance from the star until the PDR front, after which they start to increase again. We note that the 11.2/3.3 ratio decreases quickly from $\sim$ 4~$\pm$~1 to $\sim$  2.5~$\pm$~0.5 going from 45 to 55$^{\prime\prime}$  away from the star. It then tends to be relatively constant between  55--70$^{\prime\prime}$  before dropping to a minimum of $\sim$ 1.8~$\pm$~0.1 near the PDR front. In contrast, the IRAC~8.0/11.2 PAH intensity ratio displays a slower decrease moving away from the star, until it becomes relatively flat between 65--75$^{\prime\prime}$ . At the PDR front, the IRAC~8.0/11.2 PAH ratio shows a very small peak which thus corresponds to the local maximum at the PDR front as discussed above (see also Figure~\ref{ngc2023_ratios}).

\begin{center}
 {\it NGC~7023}    
\end{center}

Figure~\ref{ngc7023_ratios} shows 11.2/3.3 and IRAC~8.0/11.2 ratios in the NW PDR of NGC~7023. Note that these images have been rotated to align with the astrometry of NGC~7023 H$_{2}$ data from \cite{lem96}. Contours of the 2.124 $\mu$m H$_{2}$ line are overplotted in white to probe the PDR front. The central star is positioned beyond the bottom left edge of this image. The 11.2/3.3 intensity ratio is at a minimum all along the PDR front. Moving towards the star, this ratio rises while also showing some smaller local fluctuations on top of the overall increase with respect to the PDR front. In addition, it does not trace very well the ring structure observed in the IRAC~8.0~$\mu$m and FLITECAM~3.3~$\mu$m images. Behind the PDR front, the ratio rises although a S/N decrease limits the extent to which this can be traced reliably.

Overall, the IRAC~8.0/11.2 ratio decreases in an almost parallel layered structure with distance from the central star and does not show the ring structure as seen in the IRAC~8.0 emission. While the lowest values are found in regions sandwiching the PDR front, local maxima are apparent along the PDR front itself relative to its immediate surrounding area (around pixels (5, 20) and (13, 18)). Behind the PDR front, away from the central star, the ratio seems to rise again. 

Figure~\ref{RNe_lineprofiles} shows a crosscut across the NW PDR (see Figure~\ref{ngc7023_images} for the orientation of the cut) of the normalized fluxes of the 3.3, IRAC~8.0, and 11.2~$\mu$m emission as well as the 11.2/3.3 and IRAC~8.0/11.2 intensity ratios. Each of the emission features peak close to or at the PDR front (as traced by the peak intensity of the H$_2$ emission, at 46.3$^{\prime\prime}$  from the illuminating source). In NGC~7023, as was also found for NGC~2023, the 3.3~$\mu$m emission decreases more sharply away from the PDR front compared to the 8.0 and 11.2~$\mu$m emission. Moving towards the star, both the 3.3 and IRAC~8.0~$\mu$m intensities show an initial decrease followed by a plateau between $\sim$~30--40$^{\prime\prime}$ before decreasing again close to the star. This plateau in emission is likely due to the ring of emission seen south of the PDR front detected at both wavelengths (Figure~\ref{ngc7023_images}). The 11.2 emission shows a linear decrease in intensity as the portion of the ring structure facing the illuminating star is much fainter in this band. The 11.2/3.3 radial profile shows a decrease away from the star up to a distance of 25$^{\prime\prime}$. The ratio then increases up to a maximum at $\sim$~38$^{\prime\prime}$ due to the filaments comprising the ring structure perpendicular to the PDR front becoming increasingly luminous in the 11.2~$\mu$m. Beyond this peak, there is another decrease to a minimum located just beyond the PDR front at 50$^{\prime\prime}$, after which the ratio increases again. The IRAC~8.0/11.2 radial profile shows a different overall trend. It peaks closest to the illuminating star and shows a significant decrease away from the star down to a minimum that occurs at the same distance as the maximum in the 11.2/3.3 radial profile, corresponding to the emergence of the 11.2~$\mu$m emission within the ring structure. This again can be attributed to the steadily increasing prominence of the ring structure in the 11.2~$\mu$m emission relative to the other bands which show a more uniform emission distribution here. With increasing distance from the central star, the IRAC~8.0/11.2 ratio increases again until $\sim$~50$^{\prime\prime}$  after which it drops. 

We note that while the relative trends in 11.2/3.3 and IRAC~8.0/11.2 ratios are quite similar between this study and \cite{cro16}, the absolute values of the ratios are not in agreement. This arises due to the following differences. 
First, \cite{cro16} used the FORCAST 11.1 $\mu$m broadband filter with a FWHM of 0.95~$\mu$m. These authors estimated that the continuum emission contributes 20\% to the observed emission in the FORCAST filter and, as a consequence of this low contribution, represents the 11.2~\mum\, PAH flux by the observed FORCAST 11.1 $\mu$m flux. In contrast, we measure the integrated 11.2~$\mu$m band strength from the IRS~SH cube (Section \ref{spit}). The PAH emission contributes 80\% of the observed emission at 11.2~$\mu$m at the PDR front. Due to the FORCAST filter width, this results in a lower (integrated) PAH contribution.  Indeed, we find that the integrated flux accounts for 58\% of the PDR front emission within the FORCAST bandwidth decreasing to 35\% at 20$^{\prime\prime}$  from the illuminating star. 
In addition, we note that the portion of the ring-like structure facing the illuminating source is not clearly visible in our 11.2~$\mu$m PAH map (extracted from IRS~SH spectra) whereas it is clearly evident in the FORCAST 11.2~$\mu$m image presented in Figure~1 of \cite{cro16}. We attribute this to our subtraction of the dust continuum emission from the IRS~SH spectra to isolate the 11.2~$\mu$m PAH emission. Summarizing, different methods of representing the 11.2~$\mu$m PAH emission strength and the subtraction (or not) of the underlying dust continuum give rise to differences in the 11.2~$\mu$m PAH strength between this work and \citet{cro16}. 
Second, the absolute surface brightness of the IRAC~8.0~$\mu$m observations used in each study differs. Specifically, the surface brightness at the NW PDR of the archival IRAC~8.0~$\mu$m observation used here is $\sim$~3--4 times brighter than in the IRAC~8.0~$\mu$m observation used in \cite{cro16}. The origin of this discrepancy is unknown and so we elected to use the archival IRAC~8.0~$\mu$m observation. As a consequence of these differences with the 11.2 and IRAC~8.0 measurements, our IRAC~8.0/11.2 ratios are a factor of 7 higher than those reported in \cite{cro16}. Despite these discrepancies, we emphasize that the relative trends found in both studies are consistent.

\begin{figure}[t]
\begin{center}
\includegraphics[clip,trim =0.5cm 0.2cm 1.5cm 0.2cm,width=8.5cm]{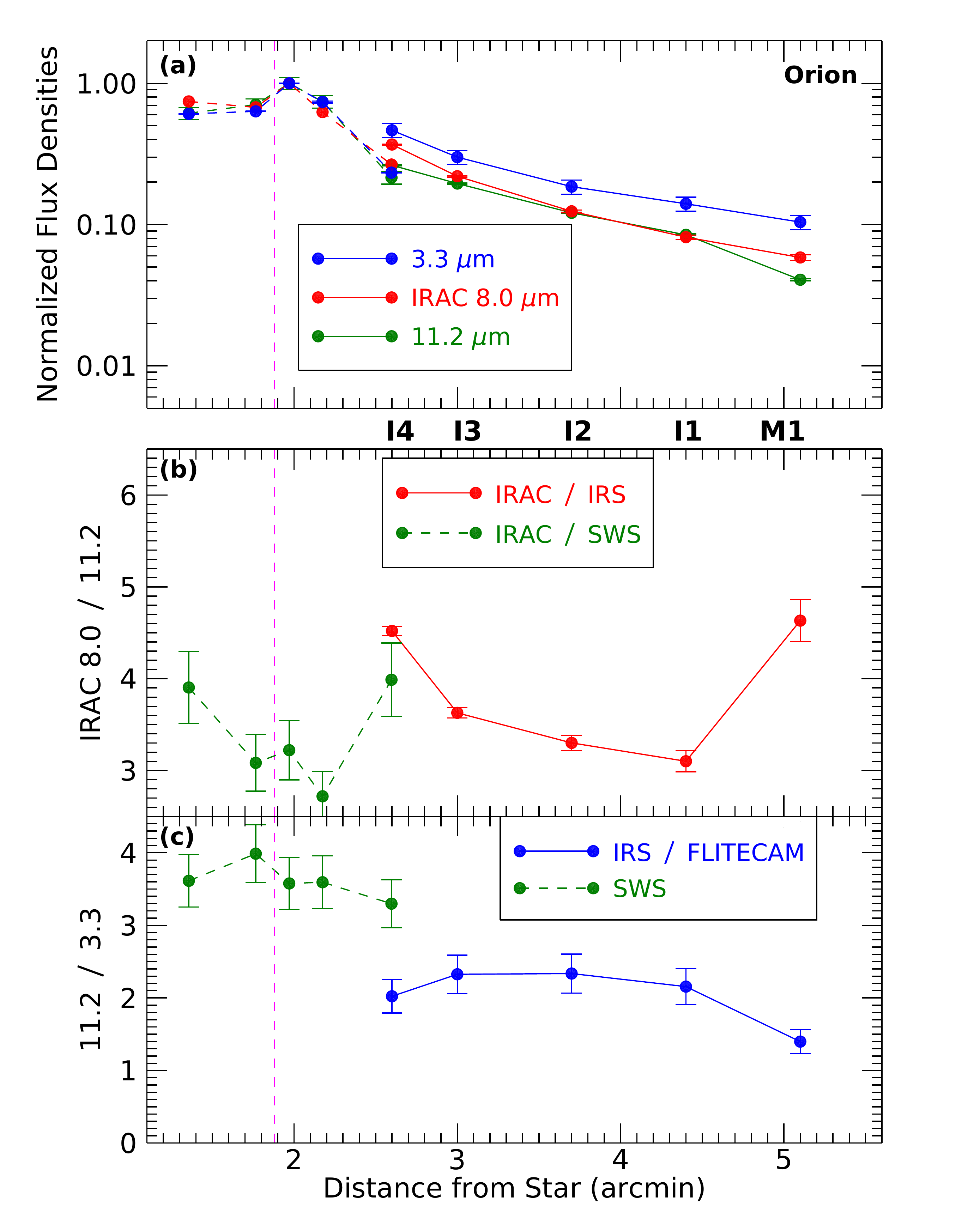}
\end{center}
\caption{Normalized radial profiles of the 3.3, IRAC~8.0, and 11.2~$\mu$m intensities (a) and radial profiles of the IRAC~8.0/11.2 (b) and the 11.2/3.3 (c) ratios as a function of distance to the illuminating source in the Orion Nebula (see Figure~\ref{Orion_images}). Radial profiles including 3.3 and 11.2~$\mu$m intensities and the associated ratios derived in each of the five ISO-SWS pointings across the Orion Bar are shown here using dashed lines. The projected distance from $\theta^{1}$ Ori C to the ionization front of the Orion Bar (shown in magenta) is taken from [OI] 6300~\AA \ measurements of \cite{sal16}. Note that increasing 11.2/3.3 corresponds to increasing PAH size and increasing IRAC~8.0/11.2 corresponds to increasing ionization. The associated uncertainties are indicated as error bars of the same color as the respective data.}
\label{orion_lineprofiles}
\end{figure}

\begin{center}
 {\it Orion}    
\end{center}
Figure~\ref{orion_lineprofiles} shows the normalized flux densities of the 3.3, IRAC~8.0, and 11.2~$\mu$m PAH emission as well as the 11.2/3.3 and IRAC~8.0/11.2 ratios as a function of distance from the illuminating source for each IRS pointing. 
Each of the three emission features are brightest near the Bar and decrease steeply out to the M1 pointing. Note that all three of these features vary in lockstep across the Orion Bar. Subsequently, moving outwards towards the Veil, the IRAC~8.0 and 11.2~$\mu$m exhibit a slower rate of decrease with distance reaching a minimum at the M1 pointing. 
Differences in the relative decline of the band intensity are better probed with their intensity ratios. Across the I1--I4 positions, the 11.2/3.3 ratio remains constant within the associated uncertainties at 2.3~$\pm$~0.2 and decreases at the M1 position to a ratio of 1.4~$\pm$~0.1 due to the larger relative decrease in the 11.2~$\mu$m band. 
In contrast, the IRAC~8.0/11.2 intensity is high at the I4 position and decreases down to a minimum at the I1 position. This is followed by an increase back to a maximum at the M1 position, again due to the 11.2~$\mu$m emission showing the sharpest relative decrease at this pointing. 

\section{Discussion}
\label{discussion}

Relative intensity variations of the 3.3 to 11.2~$\mu$m bands trace the average size of the observed PAH population \citep[e.g.][]{sch93,ric12}. Indeed, upon absorption of a UV photon of a given energy, a smaller PAH molecule, with less vibrational modes in comparison to a larger molecule, will obtain a higher internal temperature as more energy is being distributed per mode. Smaller PAHs will thus emit preferentially in the 3.3~$\mu$m band compared to the 11.2~$\mu$m band. Both the 3.3 and 11.2~$\mu$m emission bands originate in neutral PAH species which eliminates any dependence on the PAH charge state. Both bands also originate in CH modes and are therefore not sensitive to variations in the H/C ratio of the emitting species. In addition, these bands span the largest range in wavelength of the major PAH bands \citep[excluding the 12.7 $\mu$m, which is of mixed charge state, i.e.][]{pee12, pee17, sha15, sha16}, maximizing the spectral sensitivity to internal temperature. 

\cite{ric12} employed the NASA Ames PAH IR spectroscopic database \citep{bau10} to calculate the intrinsic emission spectrum of PAHs with different sizes for an average photon energy of 6 and 9~eV. These authors find i) a clear inverse-relationship between the 3.3/11.2 PAH band ratio and PAH size for PAHs of the coronene and ovalene family and ii) that a higher absorbed average photon energy will increase the 3.3/11.2 ratio for a given PAH species (their Figure 16). Hence, comparing the observed 3.3/11.2 PAH intensity ratios with these calculations can then provide an estimate for the expected size distribution of the PAH population.

In contrast, the IRAC~8.0/11.2 ratio traces the ionization balance of the PAH population. Indeed, the IRAC~8.0~$\mu$m channel covers the 7.7 and 8.6~$\mu$m PAH emission bands which are due to ionized PAH molecules while the 11.2~$\mu$m emission band arises from neutral PAH species. Hence, this ratio yields another probe of the interaction between FUV radiation and the PAH population.

\cite{cro16} investigated the size and charge distribution of the PAH population in NGC~7023. These authors followed the approach by \cite{ric12} to determine the average size of the PAH population based on the 11.2/3.3 PAH intensity ratio using an average absorbed photon energy of 6.5~eV (their Figure 5). They report that the average PAH size reaches a minimum of 50~carbon atoms at the PDR front and thus increases away from the PDR front towards both the molecular cloud and the illuminating star reaching sizes up to 70~carbon atoms. In addition, these authors report that the charge distribution becomes increasingly dominated by ionized PAHs with decreasing distance from the central star, consistent with previous results \citep[e.g.][]{job96,ver96,ber07}. 

Comparing the (relative) PAH intensity radial profiles for NGC~7023 and NGC~2023 (Figure~\ref{RNe_lineprofiles}), we report very similar trends in both RNe which are consistent with the results of \cite{cro16}: i) the intensity of each emission component increases with distance from the illuminating source and peaks at the PDR front before decreasing again into the molecular cloud, ii) the size of the PAH population reaches its minimum at the PDR front, and iii) the charge state of the PAH population decreases from the illuminating source towards the PDR front. As we probe a different configuration with our pointings towards the Orion Bar, its trends, as described in Section ~\ref{results}, are clearly distinct compared to those of the RNe. We now consider the specific properties of the size distribution and charge balance of the PAH population for each source individually. 

\subsection{Average PAH size distribution}

We derive the average PAH sizes for our sample based on the 11.2/3.3 ratios as observed with FLITECAM and IRS. We remind the reader that some uncertainty exists related to their absolute values (cf. FLITECAM--SWS comparison, Section \ref{results}) and we did not apply a correction for the PAH contribution in the FLITECAM filter. Both will affect the derived average PAH sizes. Consequently, trends in the average PAH size are trustworthy while its absolute value is indicative.  

\paragraph{NGC~7023}
To determine the average PAH size in NGC~7023, we apply the emission model shown in Figure~5 of \cite{cro16}. The minimum in the 11.2/3.3 ratio of $\sim$ 1.7~$\pm$~0.1 at the NW PDR front (Figure~\ref{ngc7023_ratios} and Figure~\ref{RNe_lineprofiles}) corresponds to an average PAH size of $\sim$ 50~$\pm$~2~carbon atoms. Closer to the illuminating star, the 11.2/3.3 ratio rises up to 2.9~$\pm$~0.2, corresponding to PAHs with $\sim$ 65~$\pm$~5~carbon atoms. This is followed by a decrease to a ratio of 1.8~$\pm$~0.1 or PAHs with $\sim$ 50~$\pm$~2~carbon atoms before rising again into the cavity surrounding the star. As detailed in Section \ref{results}, the trends found in the 11.2/3.3 and consequently the average PAH size for the NW region of NGC~7023 are overall similar to those reported by \cite{cro16} while absolute values differ due to the different methods used to measure the 11.2~$\mu$m emission. 

\paragraph{NGC~2023}
We use the Kurucz stellar model used by \cite{and15} and the emission model of \citet[][which takes the absorption cross--section into account]{cro16} to obtain an average photon energy of 7.3~eV. Using the dependence of the 11.2/3.3 PAH ratio on size as determined by \citet[][their Figure 16]{ric12}, we obtain an average PAH size of 75~$\pm$~5~carbon atoms for a minimum value of the 11.2/3.3 ratio of $\sim$ 1.8~$\pm$~0.1 at the S and SSE Ridges (Figures~\ref{ngc2023_ratios} and~\ref{RNe_lineprofiles}). Moving towards the star, the 11.2/3.3 ratio rises up to $\sim$ 4~$\pm$~1 corresponding to a size of $\sim$~110~$\pm$~20~carbon atoms.

\paragraph{Orion}
By using the emission model of \cite{cro16} with a Kurucz stellar model of T$_{eff}$~=~39000~K and log(g)~=~4 for $\theta^{1}$ Ori C \citep{kur93}, we obtain an average photon energy of 8.1~eV\footnote{We again note that \cite{ode17} have suggested that the primary ionization source beyond the Orion Bar is $\theta^{2}$ Ori A, a O9.5V type star with T$_{eff}$~=~34600 K. This corresponds to an average photon energy of 7.9~eV, which does not significantly change our size estimates.}. The 11.2/3.3 ratios in all regions probed SE of the Orion Bar vary between approximately 2.3~$\pm$~0.2 to 1.4~$\pm$~0.1\footnote{We note that these ratios and the subsequent PAH size derivations are highly influenced by the large multiplication factor used in Orion FLITECAM calibration. Hence, we emphasize the absolute values should be considered highly uncertain while the relative variations are trustworthy. }. This corresponds to an average PAH size ranging from 70~$\pm$~5 to 85~$\pm$~5~carbon atoms based on the 11.2/3.3 PAH ratio and size relationship as determined by \citet{ric12}. \\

Considering these three sources, we thus find that i) the average PAH sizes in NGC~7023 are the smallest, with its maximum average PAH size being similar to the minimum average PAH size in both NGC~2023 and SE of the Orion Bar, and ii) the range of average PAH size across the FOV is largest towards NGC~2023 (Table~\ref{tab:physcond}). Comparing these results with the physical conditions of the sources (Table~\ref{tab:physcond}), we note that for the RNe, the average PAH size increases when the intensity of the radiation field, as measured by G$_0$, increases.

\begin{table*}[t]
    \caption{Physical conditions and average PAH size for our sample.}
    \label{tab:physcond}
    \centering
    \begin{tabular}{llccccccl}
     Source      &  \multicolumn{1}{c}{Coordinates$^1$} & G$_0$ $^2$ & n$_e$ $^3$ & T$_{gas}$ & $\gamma$ $^4$ & \multicolumn{2}{c}{Average PAH size} & Refs\\
                 & \multicolumn{1}{c}{(J2000)} & & (cm$^{-3}$)& (K) & ($\times$ 10$^2$) & minimum & maximum \\
     \hline\\[-10pt]
     NGC~7023    &  (21:01:33.3; +68:10:17.60) & 2600 & 0.64 & 250 & 20 & 50~$\pm$~2 &  65~$\pm$~5 & 1 \\
     NGC~2023    &  (05:41:40.39; -02:16:02.48) $^5$ & 1.5 $\times$ 10$^4$ & 16  & 750 & 8 & &  & 2 \\
                &  (05:41:37.99; -02:16:35.26) $^6$ &  4000  &   &   &   &  75~$\pm$~5  & 110~$\pm$~20 & 2  \\
     Orion Bar & (05:35:19.9; -05:25:09.82) $^7$  &  4 $\times$ 10$^4$ & 8   & 500 & 88 &  70~$\pm$~5 &  85~$\pm$~5 & 3, 4\\
     \hline \\ [-10pt]
    \end{tabular}
    
    $^1$ Position at which the physical conditions were derived; 
    $^2$ G$_0$ is the integrated 6--13.6~eV radiation flux in units of the Habing field = $1.6 \times 10^{-3 }\, {\rm erg \, cm^{-2}\,  s^{-1}}$; 
    $^3$ electron density in the PDR (cm$^{-3}$) estimated by $n_e \approx (C/H)n_H = 1.6 \times 10^{-4}\, n_H$ using the Carbon abundance of \citet{sofia04}; 
    $^4$ Ionization parameter $\gamma = (G_0/n_e) (T_{gas}/1000{\rm K})^{0.5}$; 
    $^5$ corresponds to the peak IR emission located outside our IRS FOV;
    $^6$ corresponds to the S ridge position;
    $^7$ Orion Bar position 4, which corresponds to the ISO-SWS D5 position.
    
    References: 1. \cite{cho88}; 2. \cite{ste97}; 3. \cite{tau94}; 4. \cite{gal08}. 

\end{table*}

\subsection{PAH charge distribution}
\paragraph{NGC~7023}

As discussed in Section \ref{results}, the values of the IRAC~8.0/11.2 ratio are significantly higher than those derived by \cite{cro16}. Yet the relative ionization of the PAH population in NGC~7023 shows the same overall decreasing trend with distance from the illuminating star as well as a local peak along the PDR front as reported by \cite{cro16}.

\begin{figure*}[!htbp]
\begin{center}
\resizebox{\hsize}{!}{%
\includegraphics[clip,trim =.5cm .5cm .4cm 0cm,width=7.cm, angle =33.16]{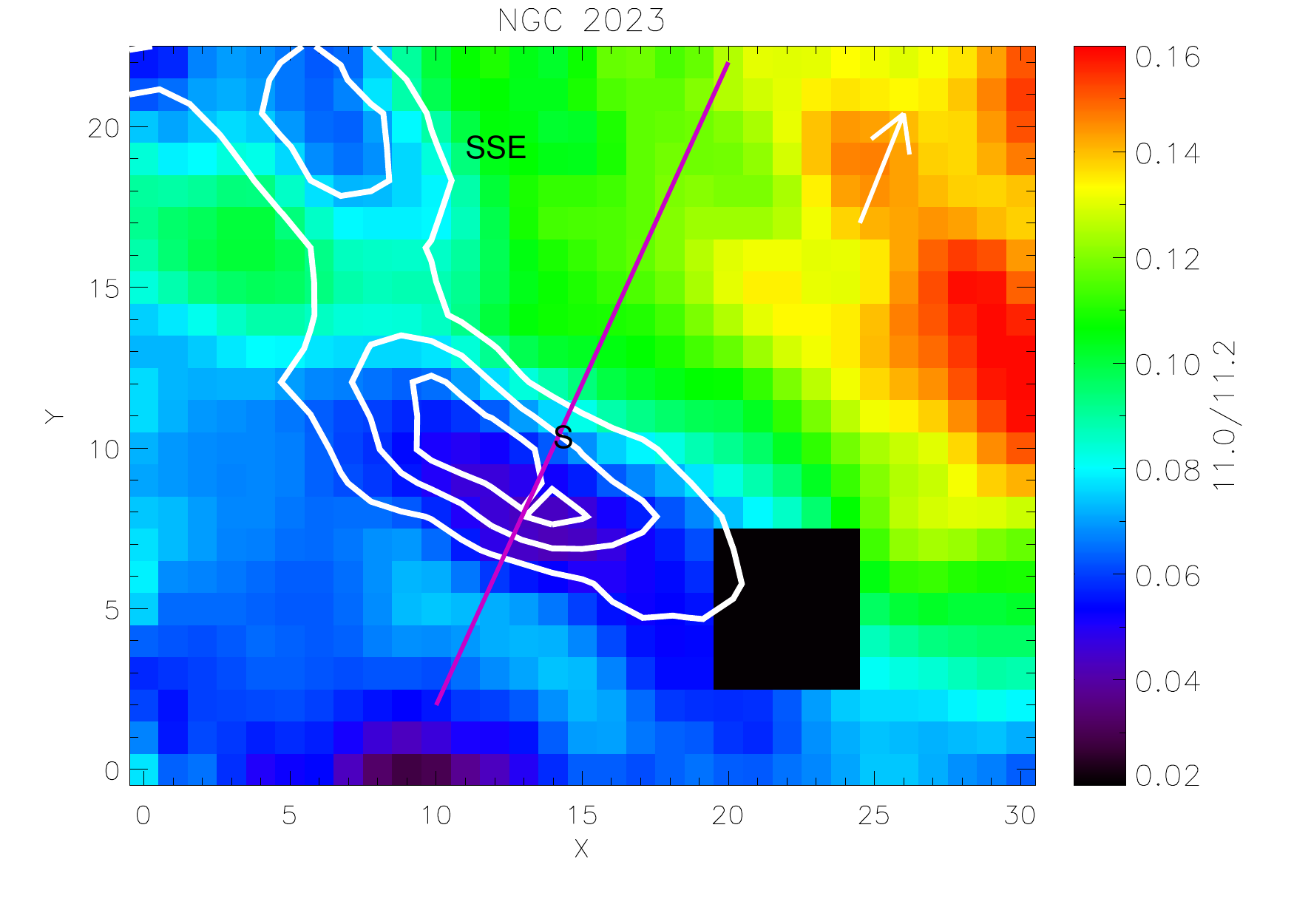}
\includegraphics[clip,trim =.5cm .5cm .5cm 0cm,width=8.5cm ]{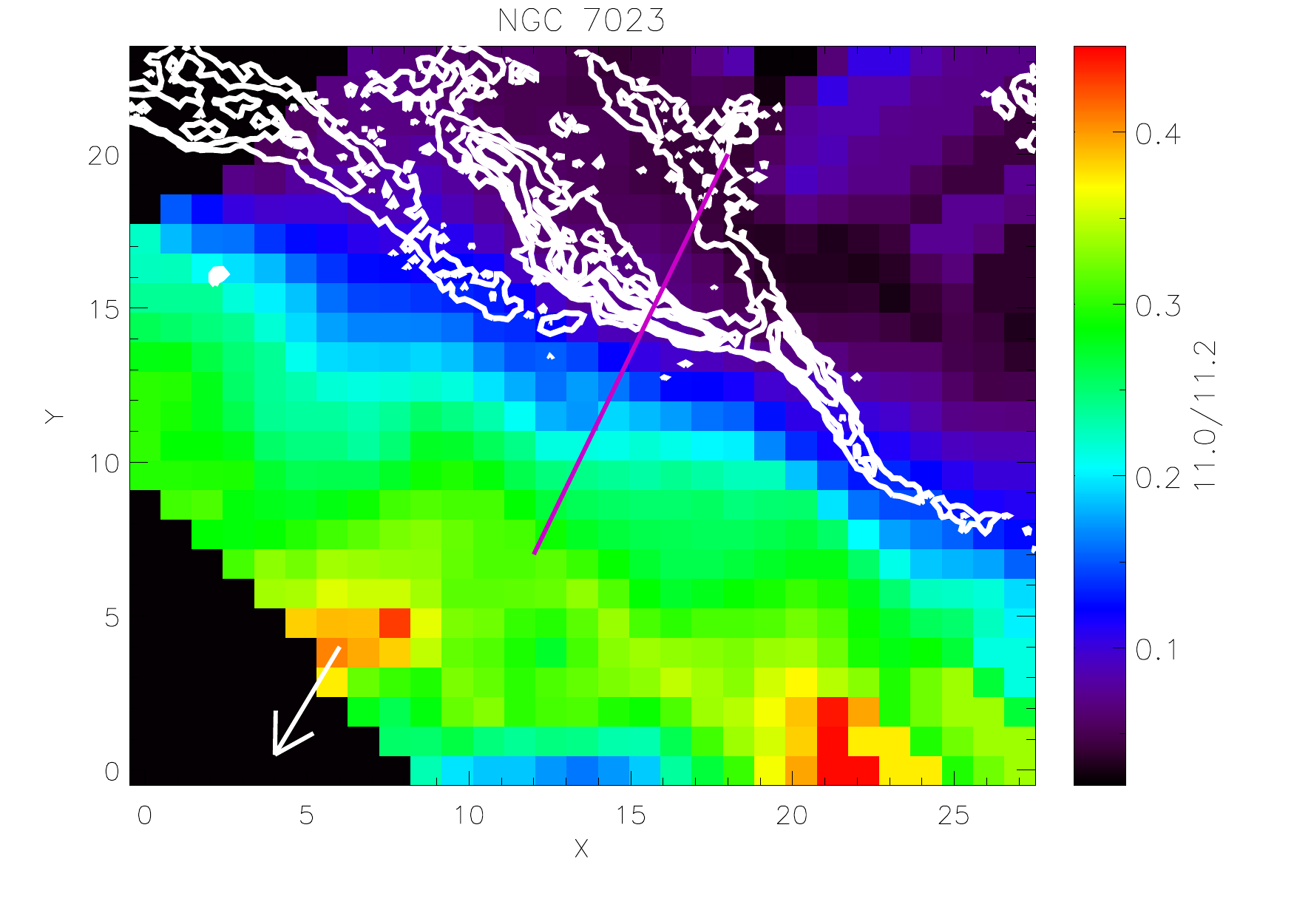}}
\resizebox{\hsize}{!}{%
\includegraphics[clip,trim =.8cm 1.7cm 1cm 2.cm,width=8.5cm]{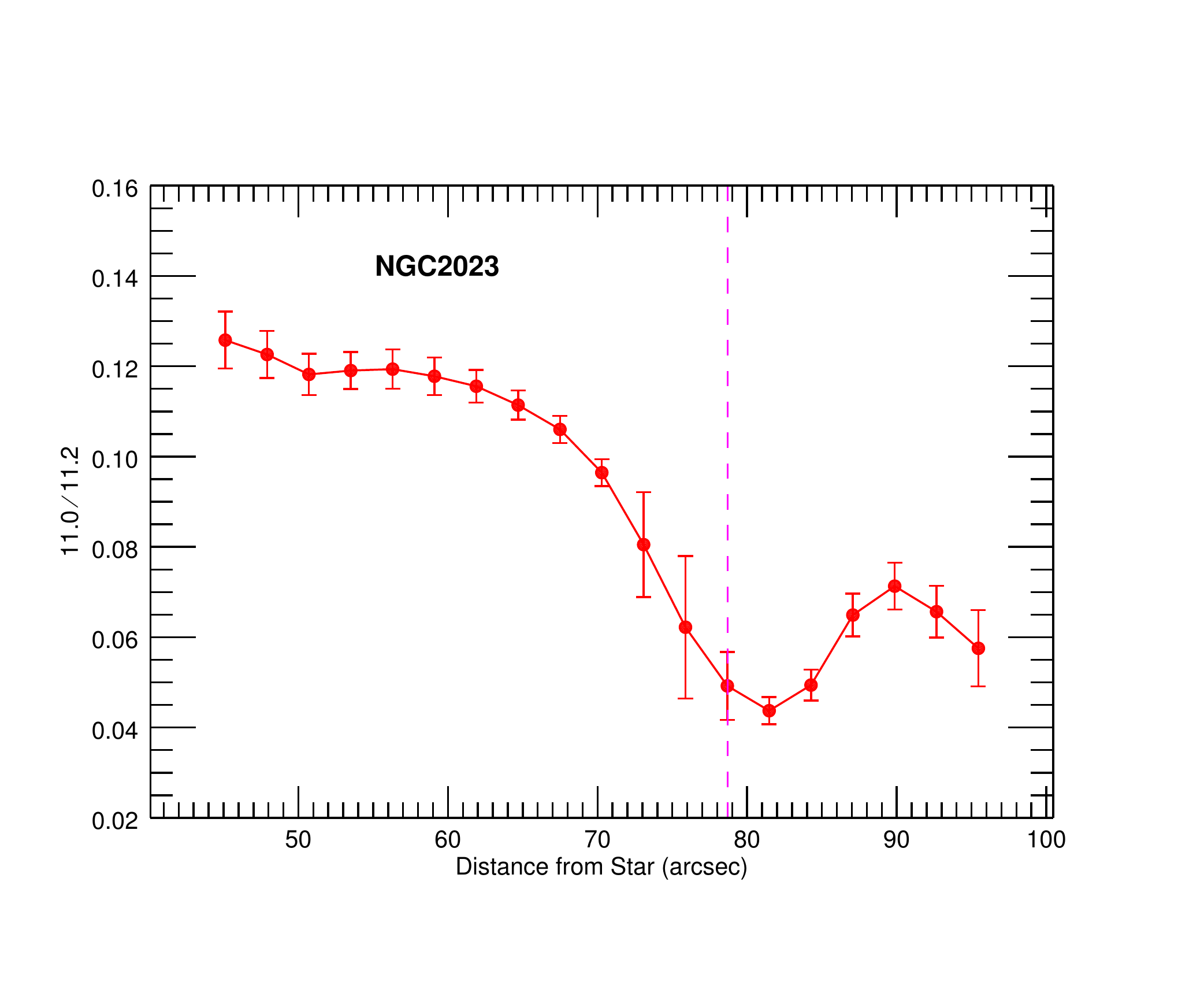}
\includegraphics[clip,trim =.8cm 1.7cm 1cm 2.cm,width=8.5cm]{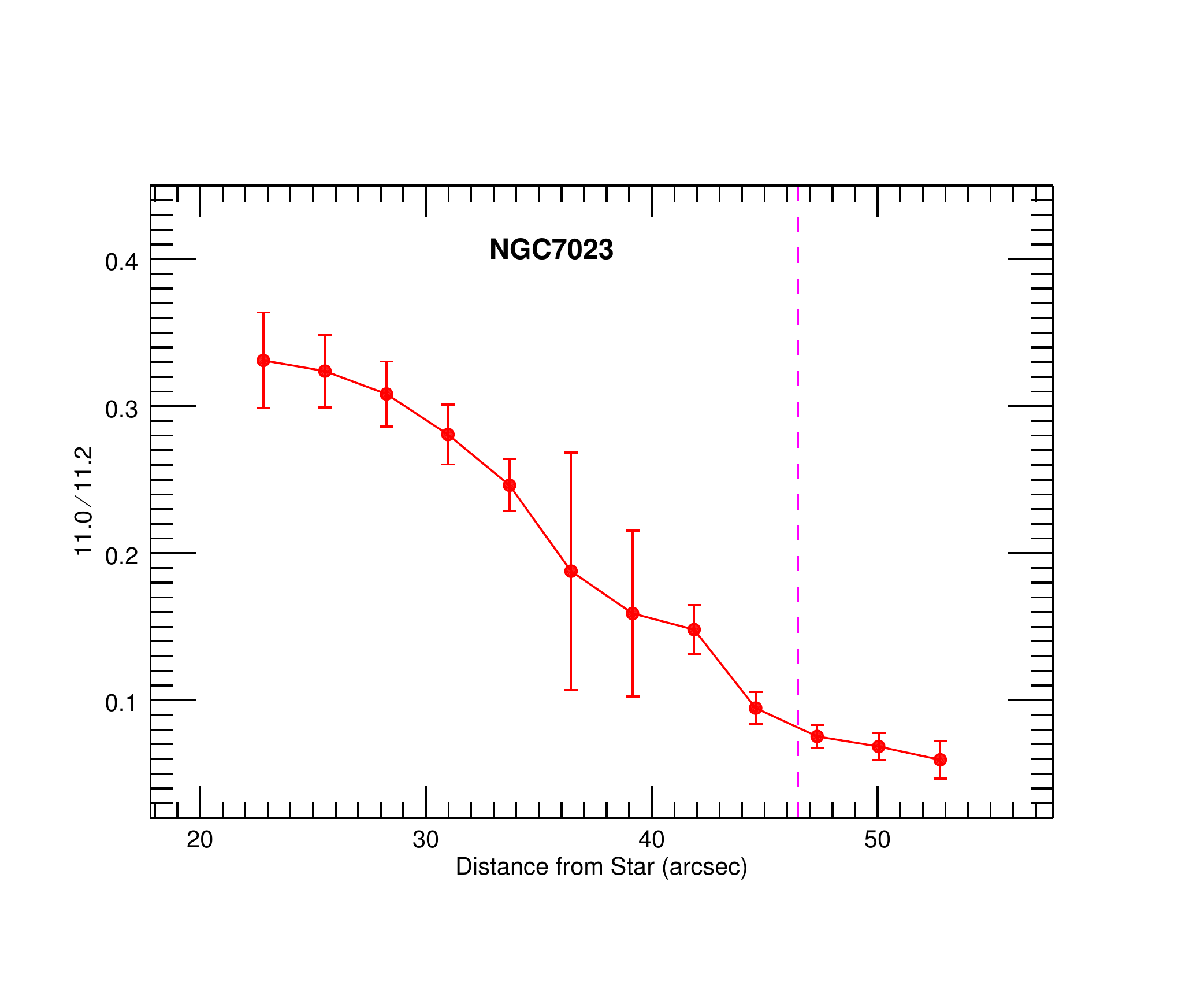}}
\end{center}
\caption{ The 11.0/11.2 PAH intensity ratio for the south FOV in NGC~2023 (left) and for the northwest NGC~7023 FOV (right). See Figures \ref{ngc2023_ratios} and \ref{ngc7023_ratios} for details on the contours shown for NGC~2023 and NGC~7023 respectively. Pixels below a 3~$\sigma$ detection or outside of the IRS~SH aperture are set to zero (shown here in black). The direction to the illuminating source is indicated by white arrows. The magenta line (top) indicates the cut used for the radial profiles. Radial profiles of the PAH ionization balance as traced by the 11.0/11.2 PAH ratio as a function of distance to the central star are shown in the lower panels. The PDR front is traced by the 12.3 $\mu$m 0--0 S(2) H$_{2}$ emission line in NGC~2023 and the 2.124 $\mu$m 1--0 S(1) H$_{2}$ emission line in NGC~7023 (magenta, bottom). The associated uncertainties are indicated as error bars.}
\label{fig_rne_ionizationbalance}
\end{figure*}

Another tracer for the PAH ionization balance is the 11.0/11.2 PAH ratio \citep[e.g.][]{hud99, ros11, sha16, pee17} as the 11.0 and 11.2 PAH emission are due to solo-CH out-of-plane bending modes in respectively cationic and neutral PAHs \citep[e.g.][]{hud99, hon01, bau08}. A crosscut of the 11.0/11.2 PAH ratio in NGC~7023 shows a clear decrease with distance from the illuminating source (see Figure  \ref{fig_rne_ionizationbalance}). This is fairly consistent with the crosscut of the IRAC~8.0/11.2 ratio in NGC~7023 shown in Figure \ref{RNe_lineprofiles}. 

\paragraph{NGC~2023}
Regarding the degree of ionization, we noticed the existence of a local maximum in the S ridge whereas no similar peak is seen at the SSE ridge (Figure~\ref{ngc2023_ratios}). The Two Micron All Sky Survey \citep[2MASS,][]{skr06} observations of NGC~2023 detect a stellar source embedded within the south ridge located $\sim$ 2.5$^{\prime\prime}$  to the west of our line cut through this region (see Figure \ref{ngc2023_images}), which possibly gives rise to this increased PAH ionization balance.

There are clear difference between the IRAC~8.0/11.2 ratio and the 11.0/11.2 PAH ratio in NGC~2023 (see Figure  \ref{fig_rne_ionizationbalance}). The 11.0/11.2 PAH ratio does not show the local maximum at the PDR front seen in the IRAC~8.0/11.2, but a local minimum just behind this point. Additionally, the 11.0/11.2 PAH ratio profile may reflect the bowl-shaped geometry of the cavity, where the relative ionization decrease is largest at the edge-on PDR front. There is a slight rise in the 11.0/11.2 PAH ratio moving into the molecular cloud, but it is nowhere near as pronounced as the rise found in the IRAC~8.0/11.2 ratio of NGC~2023 (see Figure \ref{RNe_lineprofiles}). The discrepancy between the IRAC~8.0/11.2 ratio and the 11.0/11.2 PAH ratio seen beyond the PDR front suggests that the increasing continuum contribution of the IRAC~8.0~$\mu$m makes the IRAC~8.0/11.2 ratio a less reliable tracer of PAH ionization into the molecular cloud (e.g. Appendix \ref{appa}). 

\begin{figure}[t]
\begin{center}
\includegraphics[clip,trim =0.25cm 0.2cm 1.5cm 0.2cm,width=8.5cm]{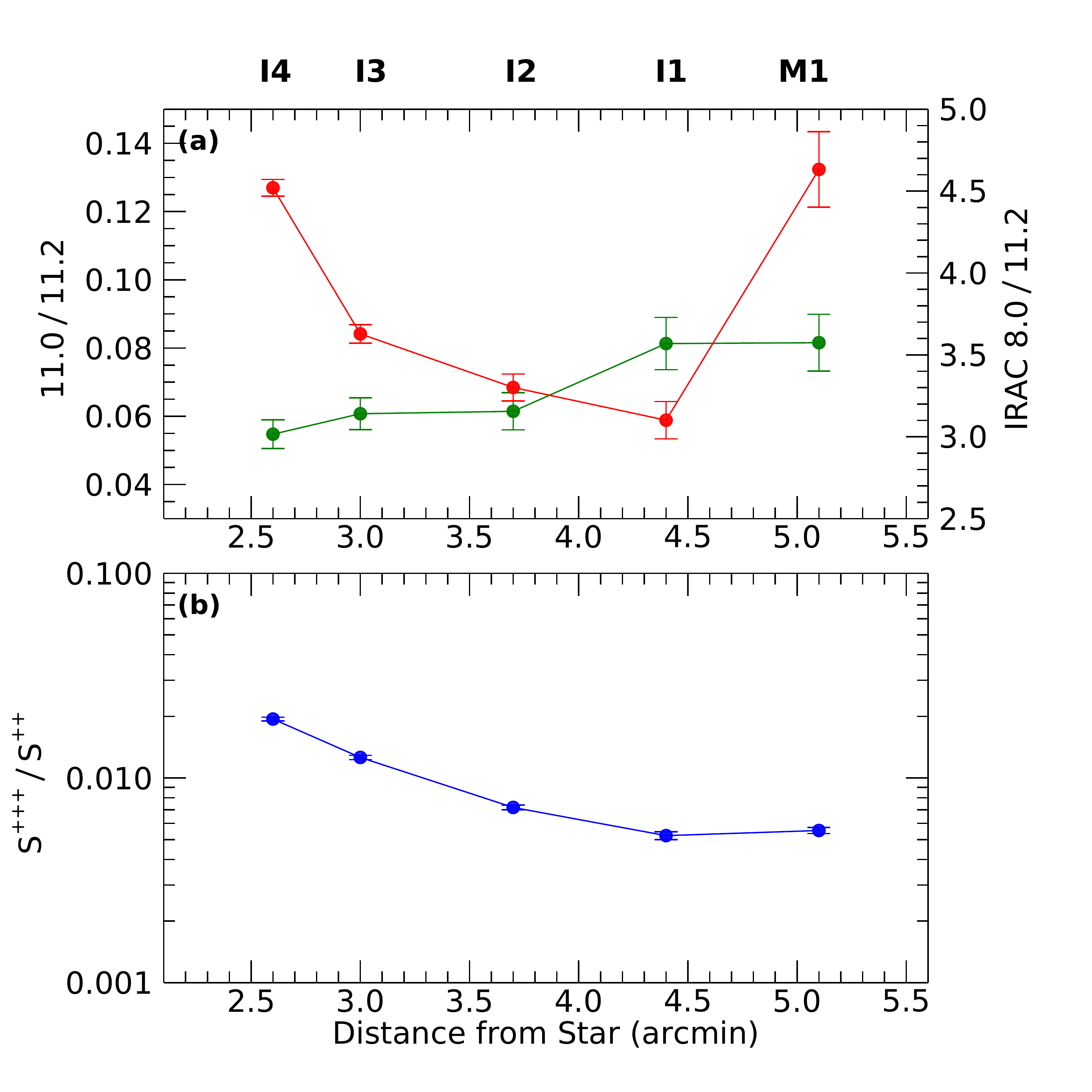}
\end{center}
\caption{The Orion Nebula radial profiles of the PAH ionization balance as traced by the IRAC~8.0/11.2 ratio (red, this work, top) and the 11.0/11.2 PAH ratio \citep[green,][top]{boe12}, and of the hardness of the radiation field as traced by the S$^{+++}$/$S^{++}$ ratio \citep[blue,][bottom]{rub11} as a function of distance to the illuminating source. The associated uncertainties are indicated as error bars of the same color as the respective data. }
\label{fig_orion_ionizationbalance}
\end{figure}

\paragraph{Orion}

The radial profile of the PAH ionization as traced by the IRAC~8.0/11.2 ratio does not resemble that of the 11.0/11.2 PAH ratio reported by \citet[][see Figure~\ref{fig_orion_ionizationbalance}]{boe12}. This discrepancy may arise from, amongst others, a possible changing PAH contribution to the IRAC~8.0~$\mu$m band and/or from PAH dehydrogenation. As discussed in Section~\ref{Spitzer}, the former cannot be the main driver of the observed variation in the IRAC~8.0/11.2 ratio. In contrast, PAH dehydrogenation will influence the 11.0, 11.2, and 8.6~$\mu$m band differently as the 11.0 and 11.2~$\mu$m bands are due to solo CH groups while the 8.6~$\mu$m band probes all CH groups in large compact PAHs. We note that the IRAC~8.0~$\mu$m band comprises both the 7.7 and 8.6~$\mu$m band with the 7.7~$\mu$m band dominating. Hence, PAH dehydrogenation may cause the IRAC~8.0/11.2 and 11.0/11.2 tracers for PAH ionization to diverge. \citet{boe12} invoke PAH dehydrogenation for these observations to explain the discrepancy between the 11.0/11.2 PAH ratio and the observed change in the hardness of the radiation field as traced by the [SIV]~10.5/[SIII]~18.7 line ratio reported by \citet[][]{rub11}. 
 
 Additionally, \cite{pee17} have shown that the 6.2 and 7.7~$\mu$m bands are spatially distinct from the 8.6 and 11.0~$\mu$m bands in NGC~2023. They found that the 11.0 and 8.6~$\mu$m bands peak much closer to the illuminating source in NGC~2023 than the 6.2 and 7.7~$\mu$m bands. Although all of these bands are attributed to ionic PAH species, the spatial diversity found suggests that the 6.2/7.7~$\mu$m and 8.6/11.0~$\mu$m bands correspond to distinct groups of PAH ions. Since the 7.7~$\mu$m band will have a significantly larger contribution within IRAC~8.0~$\mu$m than 8.6~$\mu$m, we conclude that IRAC~8.0/11.2 and 11.0/11.2 cannot both be used as identical tracers of relative PAH ionization.

\subsection{Photochemical evolution of PAHs}
The observed variation of the average PAH size gives further evidence of significant ongoing photoprocessing of these molecules. Indeed, for both RNe, NGC~7023 and NGC~2023, we find that the average PAH size distribution in each source and variations in average size between the two RNe is strongly dependent on the radiation field strength, G$_{0}$. Specifically, we find that the minimum average PAH size is found at the PDR front and that the average PAH size increases upon closer approach to the illuminating stars as G$_{0}$ increases. Likewise, smaller average PAH sizes are found towards NGC~7023 (compared to NGC~2023) which has also the lowest radiation field intensity (Table~\ref{tab:physcond}). Enhanced intensity of the radiation field thus leads to increased PAH photo--processing resulting in the destruction of smaller PAH species as predicted by \cite{all96}. This, in turn, gives rise to a larger average PAH size. Together with the enhanced emission of fullerenes near the illuminating source \citep{sell10,pee12,ber12}, these results give support to the hypothesis of the formation of fullerenes from PAH molecules. Specifically, extreme levels of UV radiation promote the complete dehydrogenation of PAHs ($>$~60~carbon atoms) that subsequently fold into loose cage structures upon losing C$_{2}$ atoms until they reach stability as C$_{60}$ molecules \citep[][]{ber12, zhe14,ber15}. The interpretation of the average PAH size distribution southeast of the Orion Bar is more complicated than that of the RNe. We find no real variation in the average PAH size (within the associated uncertainties) with distance from the illuminating source, except for the M1 pointing. 
The geometrical nature of the region delineated by the Orion Bar and the southeastern Veil is much more complex than the edge--on morphology of the PDRs found in the RNe. The Bar is an edge--on, compressed shell at the edge of a bowl created by the Trapezium cluster \citep{sal16}. In our positions out to the Veil, we see a face--on PDR at the surface of the molecular cloud \citep{pab19}. In the case of an edge--on PDR, the gas is transported into the cavity by the PDR evaporation flow and hence the material closer to the star is the "same" material as farther away except that it is more thoroughly processed by UV photons. In contrast, in the Veil, we are looking face-on and hence we are integrating over the history of this processing.

\section{Conclusions}
\label{conclusion}
We present spatial maps of the 3.3 and 11.2~$\mu$m PAH intensities, the IRAC~8.0~$\mu$m emission, and the 11.2/3.3 and IRAC~8.0/11.2 emission ratios in three prominent MIR-bright sources -- the reflection nebulae NGC~2023 and NGC~7023, and the region southeast of the Orion Bar -- based upon observations from FLITECAM onboard SOFIA as well as archival Spitzer data. These emission ratio maps provide a measure of the average PAH size distribution and the relative PAH ionization respectively. 

Both RNe exhibit similar trends in these intensity ratios: i) the average PAH size increases upon approaching the illuminating source while it is at a minimum along the PDR front and ii) the relative PAH ionization increases with proximity to the illuminating source. iii) In contrast to the 11.0/11.2 PAH ratio, the IRAC~8.0/11.2 ratio is not a suitable tracer of the relative PAH ionization into the molecular cloud region because of an increasing continuum contribution to the IRAC~8.0 band.
In contrast, the region southeast of the Orion Bar exhibits little to no variation in the average PAH size and a peculiar radial profile in the relative PAH ionization: a decrease away from the illuminating source up to 4.5$^{\prime}$, followed by a increase at $\sim$~5$^{\prime}$. This radial profile does not resemble that of the 11.0/11.2 PAH ratio.
 
We derive an average PAH size of 75~carbon atoms at the PDR front up to 110~carbon atoms nearest to the illuminating source for NGC~2023; 50~carbon atoms at the PDR front, which increases up to 65~carbon atoms near the cavity region surrounding HD~200775 for NGC~7023; and a range of 70 to 85~carbon atoms for the southeast Orion region. For the RNe, larger average PAH sizes correlate with an increase in G$_{0}$. In addition, the average PAH size in NGC~7023 is lower at all points in comparison to NGC~2023, whereas NGC~2023 has the widest range of average PAH sizes in the FOV considered. These results support the interpretation of the rise in the UV field intensity with proximity to the illuminating source driving the photochemical evolution of the PAH population that destroys all but the largest species and promote the formation of more stable fullerene molecules. Due the lack of variation in the average PAH size in the region southeast of the Orion Bar, combined with its complex geometry, we are unable to make any significant conclusions about the photochemical evolution of PAHs here. Future studies of these astronomical sources using space--borne telescopes with higher spatial resolution instruments such as the James Webb Space Telescope will allow these trends of PAH size and relative ionization to be refined in much more detail.

\section*{acknowledgements}
The authors thank the referee Kris Sellgren for reviewing this work and providing in-depth feedback.
We are especially thankful to Olivier Bern\'{e} for his role in the SOFIA proposal, sharing his data, as well as much insightful feedback. We thank Tom Megeath, Christiaan Boersma, Bavo Croiset, and Brian Fleming for sharing their data with us. We also thank Amy Fare for sharing her PAH emission model calculations. 
C.K. acknowledges support from an Ontario Graduate Scholarship (OGS). E.P. acknowledges support from an NSERC Discovery Grant and a SOFIA grant. Studies of interstellar PAHs in Leiden are supported through the Spinoza premie of NWO and Horizon 2020 funding awarded under the Marie Sk\l odowska-Curie action to the EUROPAH consortium (grant number 722346).

Based [in part] on observations made with the NASA-DLR Stratospheric Observatory for Infrared Astronomy (SOFIA). SOFIA is jointly operated by the Universities Space Research Association, Inc. (USRA), under NASA contract NAS2-97001, and the Deutsches SOFIA Institut (DSI) under DLR contract 50 OK 0901 to the University of Stuttgart. This work is based [in part] on observations made with the Spitzer Space Telescope, which is operated by the Jet Propulsion Laboratory, California Institute of Technology under a contract with NASA. This publication makes use of data products from the Two Micron All Sky Survey, which is a joint project of the University of Massachusetts and the Infrared Processing and Analysis Center/California Institute of Technology, funded by the National Aeronautics and Space Administration and the National Science Foundation. This work has made use of data from the European Space Agency (ESA) mission
{\it Gaia} (\url{https://www.cosmos.esa.int/gaia}), processed by the {\it Gaia}
Data Processing and Analysis Consortium (DPAC,
\url{https://www.cosmos.esa.int/web/gaia/dpac/consortium}). Funding for the DPAC
has been provided by national institutions, in particular the institutions
participating in the {\it Gaia} Multilateral Agreement. 

\appendix
\renewcommand{\thesubsection}{\Alph{subsection}}
\setcounter{figure}{0}
\renewcommand{\thefigure}{A.\arabic{figure}}

\subsection{Relative PAH contribution to the IRAC~8.0~$\mu$m emission in a PDR}
\label{appa}

The IRAC~8.0~$\mu$m emission is often used as a tracer for PAH emission \citep[e.g.][]{hog05,smi07,sto14}. Here, we estimate the relative contribution of the individual emission components contained within the IRAC~8.0~$\mu$m filter within the South NGC~2023 and the Northwest NGC~7023 IRS~SL spectral cubes \citep[e.g.][]{sto16,pee17,sto17}. The spectral components in these spectra within the IRAC~8.0~$\mu$m filter include the 7.7 and 8.6~$\mu$m emission features, the 8~$\mu$m bump and 5-10~$\mu$m plateau, both of which are attributed to emission of PAH and related species, the 6.9~$\mu$m H$_{2}$ line, and the underlying dust continuum \citep[where we follow the decomposition given by][]{pee17}. We multiply spectral maps of these emission components by the IRAC~8.0~$\mu$m filter response function and integrate over the relevant wavelength range of each feature. Subsequently, we determine the fractional contribution of each spectral component to the total IRAC~8.0~$\mu$m flux for a given spectrum. 

\begin{figure*}[!htbp]
\begin{center}
\includegraphics[clip,trim =0.5cm 0.0cm 0.5cm 2cm,width=14cm]{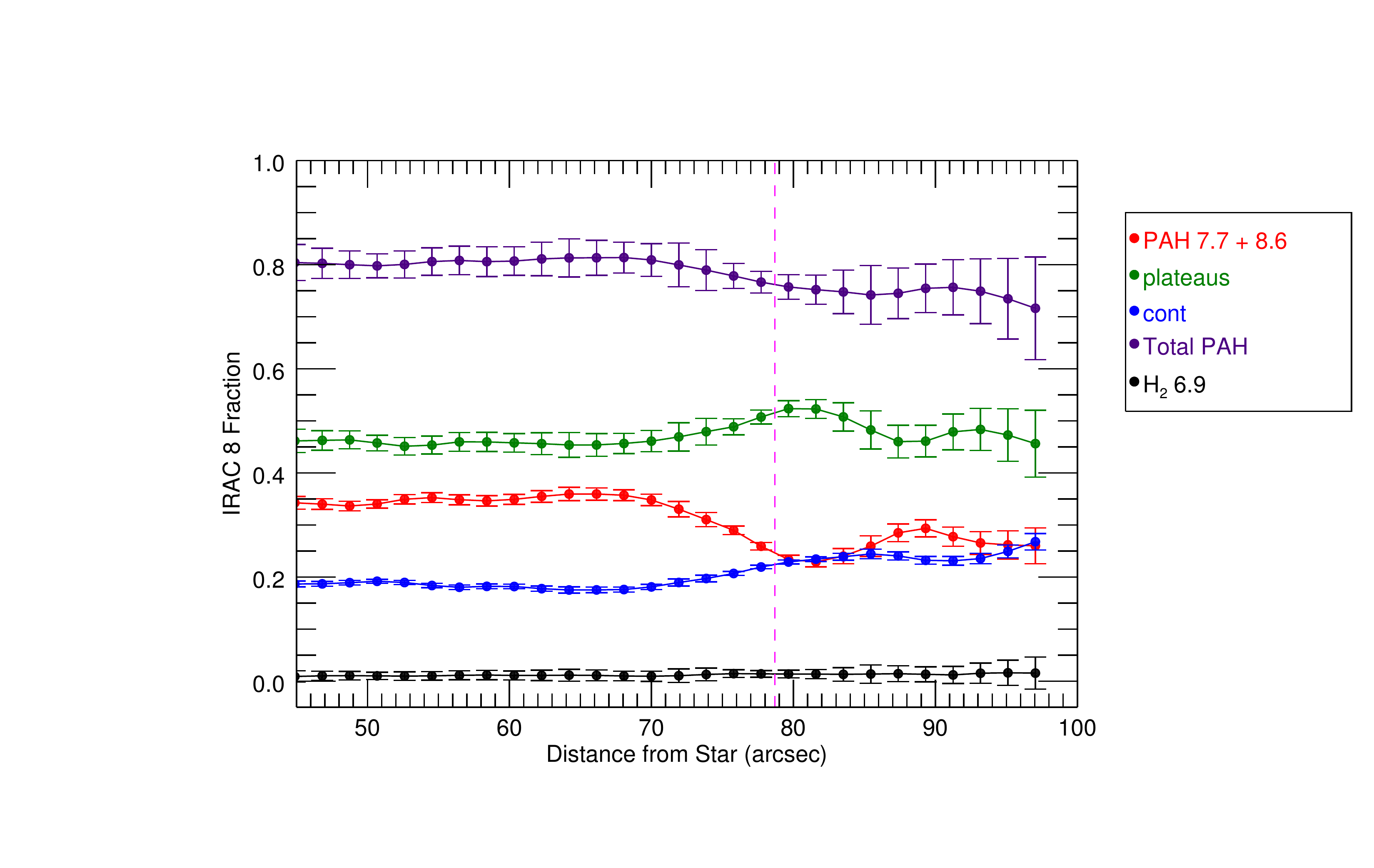}
\includegraphics[clip,trim =0.5cm 0.0cm 0.cm 2cm,width=14cm]{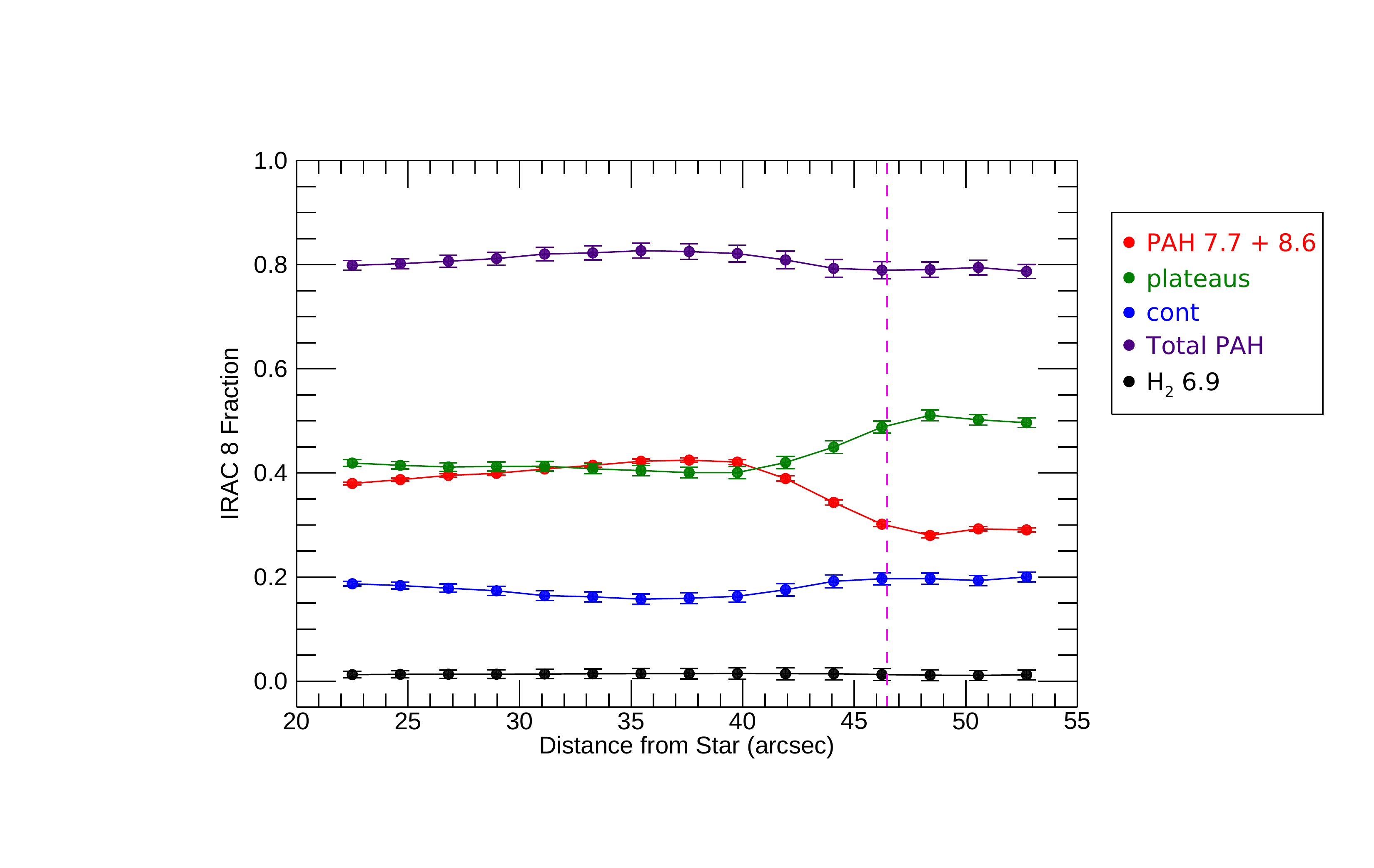}
\end{center}
\caption{Radial profile of the NGC~2023 IRS~SL South (top) and the NGC~7023 IRS~SL Northwest (bottom) data cubes taking the relative contribution of the emission features within the IRAC~8.0~$\mu$m bandpass. The PDR front is traced by the 12.3 $\mu$m 0--0 S(2) H$_{2}$ emission line in NGC~2023 and the 2.124 $\mu$m 1--0 S(1) H$_{2}$ emission line in NGC~7023 (shown in magenta). The associated uncertainties are indicated as error bars. The position of the NGC~2023 cross--cut is given in the top left panel Figure \ref{ngc2023_images} and the NGC~7023 cross-cut is shown in the top left panel of Figure \ref{ngc7023_images}.}
\label{IRAC8_PAH_contribution}
\end{figure*}

Figure \ref{IRAC8_PAH_contribution} presents these relative contributions to the IRAC~8.0~$\mu$m bandpass along radial cuts across the S ridge PDR front in NGC~2023 and the NW PDR front in NGC~7023 in the direction to the illuminating sources in both RNe. We find that along the radial cut, extending from inside the cavity to the PDR front, the PAH related species account for $\sim$~80\% of the emission within the IRAC~8.0~$\mu$m. Beyond the PDR front into the molecular cloud in NGC~2023, PAH related emission drops down to $\sim$~73\% of the total emission. Thus, we can effectively state that the IRAC~8.0~$\mu$m filter is a strong tracer of PAH emission within PDR environments but becomes a slightly less reliable tracer in molecular cloud dominated regions.

\bibliographystyle{apj}
\bibliography{mainbib}

\end{document}